\documentclass[reprint,prl,aps, floatfix]{revtex4-1}
\usepackage{graphicx}
\usepackage{color}
\usepackage{dcolumn}
\usepackage{bm}
\usepackage{amssymb}
\usepackage{braket}
\usepackage{color,amsmath}
\usepackage{comment}
\usepackage{gensymb}
\usepackage{soul,xcolor} 
\setstcolor{red} 
\usepackage{hhline}
\usepackage{tabularx}
\usepackage{xspace}
\usepackage{layouts}
\usepackage{lineno} 
\usepackage[hyperfootnotes=false,colorlinks=true,allcolors=blue]{hyperref}%pagebackref,
\usepackage{xr}
\newcolumntype{C}[1]{>{\centering\arraybackslash}m{#1}}
\newcolumntype{N}{@{}m{0pt}@{}}

\newcommand{\moire}{moir\'e\xspace}

\begin{document}
%\linenumbers

\title{Visualizing the microscopic origins of topology in twisted molybdenum ditelluride}

\author{Ellis Thompson$^{1*}$}
\author{Keng Tou Chu$^{1*}$}
\author{Florie Mesple$^{1*}$}
\author{Xiao-Wei Zhang$^{2*}$}
\author{Chaowei Hu$^{1}$}
\author{Yuzhou Zhao$^{1,2}$}
\author{Heonjoon Park$^{1}$}
\author{Jiaqi Cai$^{1}$}
\author{Eric Anderson$^{1}$}
\author{Kenji Watanabe$^{3}$} 
\author{Takashi Taniguchi$^{4}$}
\author{Jihui Yang$^{2}$}
\author{Jiun-Haw Chu$^{1}$}
\author{Xiaodong Xu$^{1,2}$}
\author{Ting Cao$^{2\dagger}$}
\author{Di Xiao$^{2,1\dagger}$}
\author{Matthew Yankowitz$^{1,2\dagger}$}

\affiliation{$^{1}$ Department of Physics, University of Washington, Seattle, Washington, 98195, USA}
\affiliation{$^{2}$ Department of Materials Science and Engineering\, University of Washington\, Seattle\, Washington\, 98195\, USA}
\affiliation{$^{3}$ Research Center for Functional Materials\, National Institute for Materials Science\, 1-1 Namiki\, Tsukuba 305-0044\, Japan}
\affiliation{$^{4}$ International Center for Materials Nanoarchitectonics\, National Institute for Materials Science\, 1-1 Namiki\, Tsukuba 305-0044\, Japan}
\affiliation{$^{*}$ These authors contributed equally to this work.}
\affiliation{$^{\dagger}$ $tingcao@uw.edu$ (T.C.); $dixiao@uw.edu$ (D.X.); $myank@uw.edu$ (M.Y.)}

\maketitle

\textbf{In moir\'e materials with flat electronic bands and suitable quantum geometry, strong correlations can give rise to novel topological states of matter~\cite{Sharpe2019,Serlin2020,Cai2023,Park2023,Zeng2023,Fan2023,Foutty2023,Kang2024,Lu2024}. The nontrivial band topology of twisted molybdenum ditelluride (tMoTe$_2$) --- responsible for its fractional quantum anomalous Hall (FQAH) states~\cite{Cai2023,Park2023,Zeng2023,Fan2023} --- is predicted to arise from a layer-pseudospin skyrmion lattice~\cite{Wu2018_topological,yu_giant_2020}. Tracing the layer polarization of wavefunctions within the moiré unit cell can thus offer crucial insights into the band topology. Here, we use scanning tunneling microscopy and spectroscopy (STM/S) to probe the layer-pseudospin skyrmion textures of tMoTe$_2$. We do this by simultaneously visualizing the moir\'e lattice structure and the spatial localization of its electronic states. We find that the wavefunctions associated with the topological flat bands exhibit a spatially-dependent layer polarization within the moir\'e unit cell. This is in excellent agreement with our theoretical modeling~\cite{Zhang_Polarization_2023}, thereby revealing a direct microscopic connection between the structural properties of tMoTe$_2$ and its band topology. Our work enables new pathways for engineering FQAH states with strain, as well as future STM studies of the intertwined correlated and topological states arising in gate-tunable devices.}

Moir\'e bilayers of semiconducting transition metal dichalcogenides (TMDs) have rapidly emerged as rich platforms for studying novel strongly correlated phases of matter~\cite{Wang2020,waters_flat_2020,Regan2020,Zhang2020,Li2021,Ghiotto2021,Huang2021,Anderson2023,Cai2023,Park2023,Zeng2023,Fan2023,Kang2024,Foutty2023,Li2021_2,Li2024}. Strikingly, the topological properties of these correlated states are remarkably sensitive to the precise atomic structure and materials composition of the moir\'e lattice. At a doping of one hole per moir\'e unit cell, correlated insulators can have associated Chern numbers of either 0 or $\pm$1 depending on the specific TMDs (e.g., WSe$_2$ vs. MoTe$_2$) and the precise twist angle, $\theta$, used to construct the moir\'e lattice~\cite{Wang2020,Cai2023,Park2023,Zeng2023,Fan2023,Kang2024,Foutty2023}. Furthermore, the long-sought FQAH states have so far only been seen in tMoTe$_2$ over a relatively narrow range of twist angles~\cite{Cai2023,Park2023,Zeng2023,Fan2023}, and have not been observed in analogous twisted WSe$_2$ structures. It is therefore critical to develop a more comprehensive understanding of the origin and nature of the nontrivial topology in these materials.

The variability of these topological properties can be traced down to microscopic origins~\cite{Zhang_Polarization_2023}, particularly the textures of electric polarization arising within the TMD homobilayer moir\'e unit cell. These generate a spatially-varying layer polarization of the flat-band wavefunction described by a layer-pseudospin skyrmion lattice. Charges hopping in this pseudospin skyrmion lattice acquire a geometric phase equivalent to the effect of a real-space magnetic field, giving rise to nontrivial band topology~\cite{Wu2018_topological,yu_giant_2020}. The locations of the north and south poles of the skyrmion lattice, where the flat-band wavefunctions become fully layer polarized, are determined by two competing mechanisms: out-of-plane ferroelectric dipole moments resulting from the broken inversion symmetry of the local stacking~\cite{Wang2022,Molino2023,Zhang2023_plasmons}, and strain-induced in-plane piezoelectric charges resulting from atomic relaxations in the moir\'e lattice~\cite{Duerloo2012,McGilly2020,shaffique2023,mao_lattice_2023}. As the structural properties of the moir\'e lattice are tuned by parameters such as twist angle and strain, these two polarization terms can conspire to flip the location of the north and south poles of the skyrmion lattice between different atomic stacking sites within the moir\'e unit cell, thereby changing the Chern numbers of the bands~\cite{Zhang_Polarization_2023}. 

%%%%%%%%%%%%%%%%%%%%%%%%%%%%%%%%%%%%%%%%%%%%%%%%%%%%%%%%%%%%%%%%%%%%%
\begin{figure*}
\includegraphics[width=\textwidth]{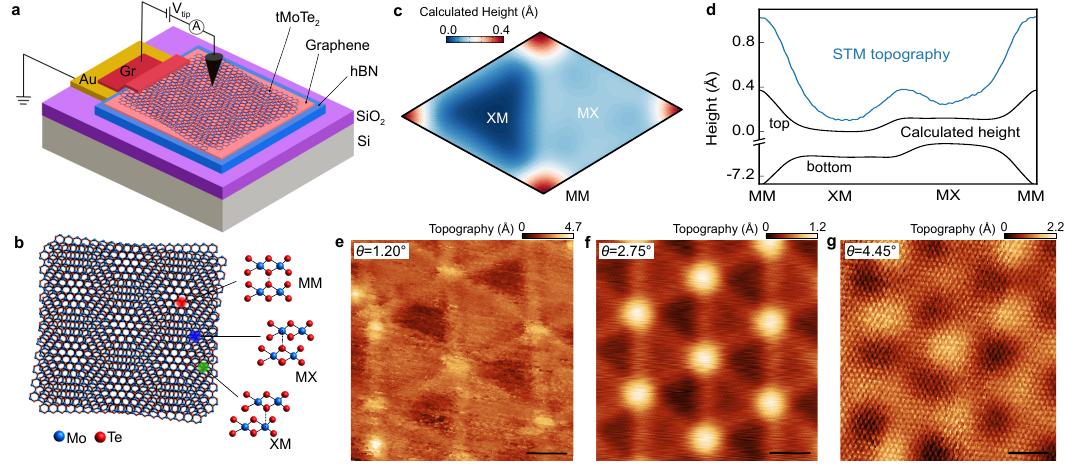} 
\caption{\textbf{Topographic characterization of tMoTe$_2$.}
\textbf{a}, Schematic of the device. tMoTe$_2$ sits atop monolayer graphene and hBN. The entire stack sits on a Si/SiO$_2$ substrate. The monolayer graphene is connected to a Au electrode by a graphite flake, and is held at ground.
\textbf{b}, Schematic illustration of a tMoTe$_2$ \moire superlattice. The interlayer stacking configuration is shown at the three high symmetry points: MM, MX, and XM.
\textbf{c}, Height of the top layer of tMoTe$_2$ with $\theta=2.88^{\circ}$ predicted by an atomic relaxation DFT calculation (see Methods for details).
\textbf{d}, Comparison between the calculated layer heights and height measured with STM topography. The black curves are line cuts of top and bottom layer heights from the calculation in \textbf{c}. The blue curve is a line cut from the topograph shown in \textbf{f}.
\textbf{e-g}, STM topographs of tMoTe$_2$ regions with varying twist angle and heterostrain. The tunneling current, $I_t$, and sample bias, $V_{bias}$, setpoints along with the structural characterization parameters are: \textbf{e}, $(I_t, V_{bias}) = (100~\text{pA},-1.8~\text{V})$, $\theta=1.20\degree$, $\varepsilon = 0.24\%$ oriented at 32$\degree$ with respect to the \moire; \textbf{f}, $(I_t, V_{bias}) = (50~\text{pA},-1.5~\text{V})$, $\theta = 2.75\degree$ , $\varepsilon_{uni} = 0.00\%$, $\varepsilon_{bi} = 0.11\%$; \textbf{g}, $(I_t, V_{bias}) = (140~\text{pA},-1.3~\text{V})$, $\theta = 4.45\degree$, $\varepsilon_{uni} = 0.71\%$ oriented at 17$\degree$ with respect to the \moire, $\varepsilon_{bi} = 0.56\%$ (see Methods for extraction of strain values, $\varepsilon$, $\varepsilon_{uni}$, and $\varepsilon_{bi}$). Scale bars are \textbf{e}, 10~nm, \textbf{f}, 5~nm, and \textbf{g}, 2~nm.
}
\label{fig:1}
\end{figure*}
%%%%%%%%%%%%%%%%%%%%%%%%%%%%%%%%%%%%%%%%%%%%%%%%%%%%%%%%%%%%%%%%%%%%%

Here, we probe the microscopic origins of topology in tMoTe$_2$ by mapping the spatial localization of valence band wavefunctions as a function of energy using scanning tunneling microscopy and spectroscopy. We fabricate high-quality van der Waals (vdW) devices consisting of tMoTe$_2$ atop monolayer graphene and hexagonal boron nitride (hBN) (Fig.~\ref{fig:1}a, Extended Data Fig.~\ref{fig:image_fab} and Methods), in which the graphene acts only as a local drain electrode for tunneling electrons~\cite{Zhang2015, pan_quantum-confined_2018, Zhang2020, tilak_moire_2023, Molino2023}. We use several spectroscopy methods to distinguish valence band states originating from the K and $\Gamma$ points of the monolayer MoTe$_2$ Brillouin zone. By studying the local density of states (LDOS) within the \moire unit cell as a function of energy, we find that the topological bands from the K points localize around only certain high-symmetry atomic stacking sites, whereas the trivial states from the $\Gamma$ point localize at all positions within the \moire unit cell. Our observations of sub-moir\'e localization of the LDOS in the flat bands are consistent with the layer-pseudospin skyrmion textures predicted by our density functional theory (DFT) calculations.

%%%%%%%%%%%%%%%%%%%%%%%%%%%%%%%%%%%%%%%%%%%%%%%%%%%%%%%%%%%%%%%%%%%%%
\begin{figure*}
\includegraphics[width=\textwidth]{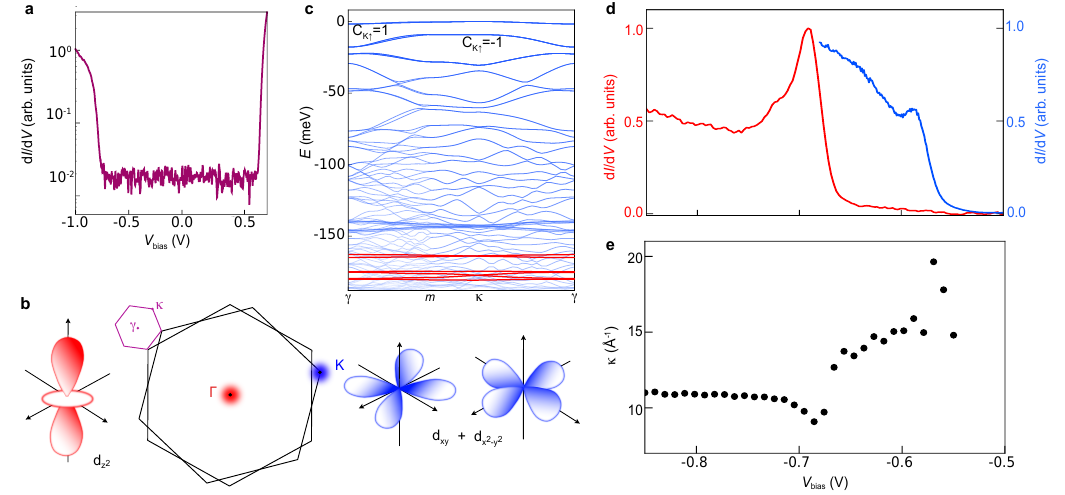} 
\caption{\textbf{Identification of spectroscopic features of tMoTe$_2$.}
\textbf{a}, Representative constant-height d$I$/d$V$ spectrum acquired over a wide range of bias in the region with a $2.75^\circ$ twist angle. The initial tunneling parameters are $(I_t, V_{bias}) = (50~\text{pA},-1.0~\text{V})$. 
\textbf{b}, Schematic of the two rotated Brillouin zones of each monolayer MoTe$_2$ (black hexagons) and their resultant moiré Brillouin zone (magenta). Cartoons schematically illustrate the orbital composition of states arising from the $\Gamma$- and K-points of the monolayer Brillouin zones.
\textbf{c}, Calculated band structure of tMoTe$_2$ with $\theta=2.88^{\circ}$ in the \moire Brillouin zone. Bands colored in blue (red) originate from the K ($\Gamma$) point of monolayer MoTe$_2$.
\textbf{d}, Representative constant-height d$I$/d$V$ spectrum with initial tunneling setpoint $(I_t, V_{bias}) = (110~\text{pA},-0.9~\text{V})$ (red), and reduced-height d$I$/d$V$ spectrum with initial tunneling setpoint $(I_t, V_{bias}) = (50~\text{pA},-1.0~\text{V})$ (blue). In the latter, the tip was moved towards the sample by a fixed distance of 0.2~nm before the bias sweep. Both curves were acquired at the MX stacking site.
\textbf{e}, Measured decay constant, $\kappa$, as a function of bias voltage (see Methods for a description of the measurement). The error in extracting the value of $\kappa$ is smaller than the marker size.} 
\label{fig:2}
\end{figure*}
%%%%%%%%%%%%%%%%%%%%%%%%%%%%%%%%%%%%%%%%%%%%%%%%%%%%%%%%%%%%%%%%%%%%% 

\medskip\noindent\textbf{Twist-angle--dependent lattice relaxations}

We first examine the moir\'e lattice structure of tMoTe$_2$ at various twist angles. When two MoTe$_2$ monolayers are twisted atop one another, three high-symmetry stacking sites can be identified based on the relative arrangement of the metal (molybdenum) and chalcogen (tellurium) atoms: MM, in which metal atoms sit atop each other; MX, in which a metal atom sits atop a chalcogen atom; and XM, in which a chalcogen atom sits atop a metal atom (Fig.~\ref{fig:1}b). Atomic relaxation effects, including in-plane strain and out-of-plane buckling, result in different heights of the top MoTe$_2$ layer at each of these three stacking sites. These effects are captured by our DFT simulations, as shown in Fig.~\ref{fig:1}c for $\theta=2.88^{\circ}$. The black curves in Fig.~\ref{fig:1}d show the calculated heights of both MoTe$_2$ layers, plotted along a cut across the moir\'e unit cell. The MX and XM stacking registries are related by $C_{2x}$ symmetry (rotation by 180$^{\circ}$ around the axis connecting nearest-neighbor MM sites).

These predictions allow us to determine the corresponding high-symmetry stacking sites in our STM topographs. Figures~\ref{fig:1}e-g show topography images of tMoTe$_2$ with twist angles of $\theta=1.20^{\circ}$, $2.75^{\circ}$, and $4.45^{\circ}$, respectively. The blue curve in Fig.~\ref{fig:1}d shows the measured height of the $\theta=2.75^{\circ}$ sample along the same cut through the \moire lattice as for the black curves. The topographic measurement is performed using a large tip bias, typically over $0.4$~V away from the bias of the valence band edge, such that the effects of the spatially varying LDOS are minimized (see Extended Data Fig.~\ref{fig:topoD1} for topographs acquired with different tunneling parameters). We find that our measurement is qualitatively consistent with the calculation, allowing us to assign the highest topography point to the MM stacking site, the second highest to MX, and the lowest to XM.

These topographs further reveal the substantial effects of lattice relaxations in tMoTe$_2$. These effects play only a subtle role at $\theta=4.45^{\circ}$, for which the MM, MX, and XM regions account for roughly comparable areas within the moir\'e unit cell. However, relaxation effects are easily visible at $\theta=2.75^{\circ}$, with the area of the MM regions shrinking relative to the triangular MX and XM domains, along with the formation of solitonic domain walls connecting the MM sites. Relaxation effects are especially profound at $\theta=1.20^{\circ}$, in which there are very large triangular MX and XM domains and comparatively small regions of MM stacking. The obvious effects of relaxations even at twist angles approaching $3^{\circ}$ are a consequence of the relative softness of the MoTe$_2$ membrane (as compared to much stiffer materials such as graphene)~\cite{shaffique2023}. Notably, large lattice relaxations play a key role in determining the topology of the flat bands in theoretical modeling~\cite{Zhang_Polarization_2023}. Understanding their microscopic properties is thus crucial for learning more about the nature of the observed FQAH states. 

%%%%%%%%%%%%%%%%%%%%%%%%%%%%%%%%%%%%%%%%%%%%%%%%%%%%%%%%%%%%%%%%%%%%%
\begin{figure*}[]
\includegraphics[width=\textwidth]{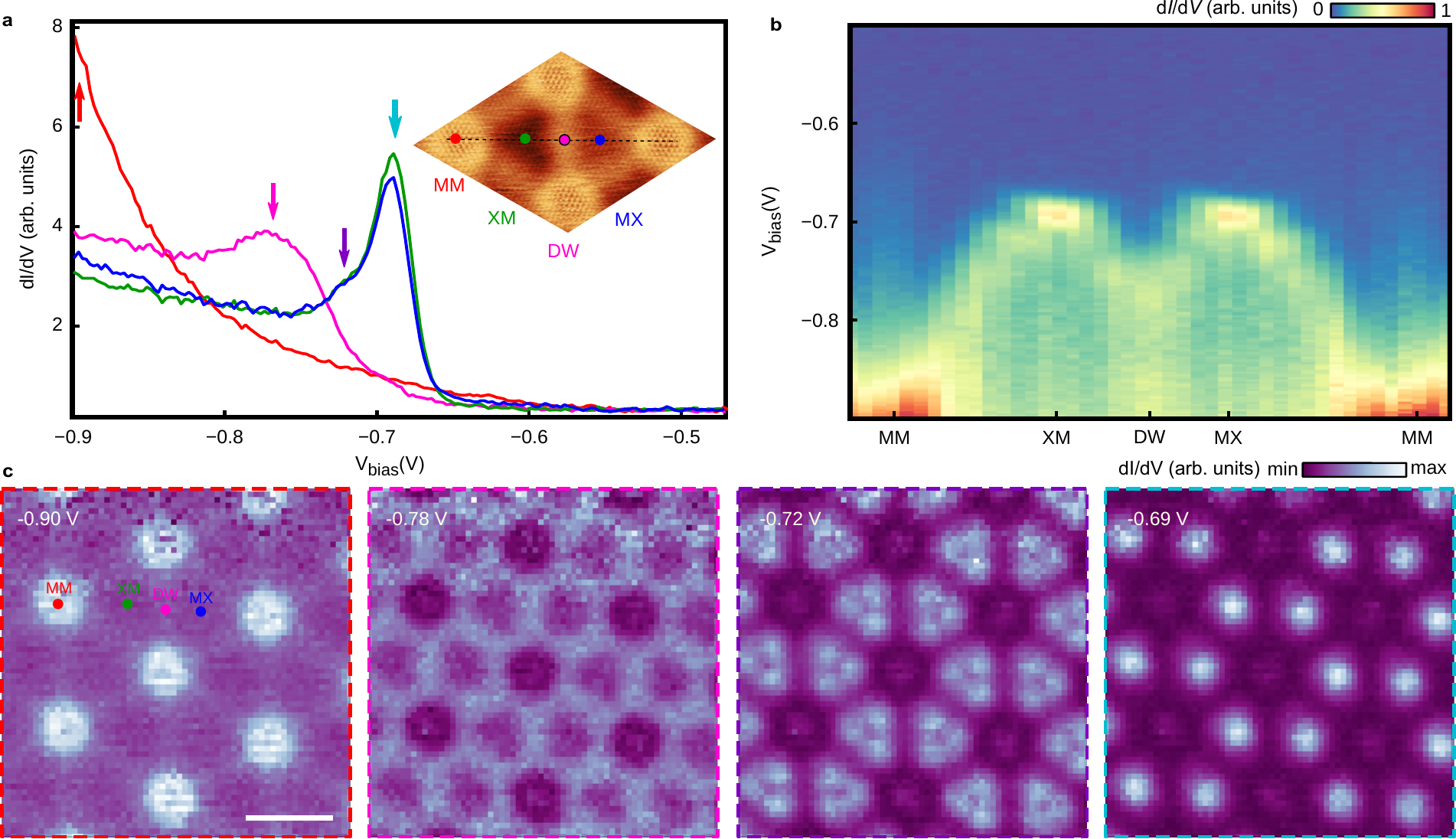} 
\caption{\textbf{Localization of $\bm{\Gamma}$-point states.} 
\textbf{a}, Constant-height d$I$/d$V$ spectra acquired at different high symmetry points (averaged across 2x2 pixels, 0.62~nm$^2$) for a tMoTe$_2$ region with $\theta=2.75^{\circ}$ (the same as in Fig.~\ref{fig:1}f). The inset shows the associated STM topograph for this region, and indicates the positions of the high symmetry points. The markers are color-coded to match the corresponding d$I$/d$V$ spectra. The dashed line is 10~nm long. 
\textbf{b}, Line cut of constant-height d$I$/d$V$ spectra acquired along the dashed path shown in the inset of \textbf{a}. 
\textbf{c}, d$I$/d$V$ maps acquired at different $V_{bias}$, as indicated by the arrows in \textbf{a}. All maps are assembled from a grid spectroscopy measurement with initial tunneling setpoint $(I_t, V_{bias}) = (110~\text{pA},-0.9~\text{V})$ (see Supplementary Video 1 for an animation of the full evolution of d$I$/d$V$ with $V_{bias}$). The scale bar is 5~nm.} 
\label{fig:3}
\end{figure*}
%%%%%%%%%%%%%%%%%%%%%%%%%%%%%%%%%%%%%%%%%%%%%%%%%%%%%%%%%%%%%%%%%%%%%

\medskip\noindent\textbf{Spectroscopic fingerprints of tMoTe$_2$}

We now turn our attention to the key spectroscopic features of tMoTe$_2$. Fig.~\ref{fig:2}a shows a representative differential conductance (d$I$/d$V$) spectrum obtained over a large range of sample bias, $V_{bias}$. The broad region of very small differential conductance surrounding $V_{bias}=0$ corresponds to the semiconducting band gap of tMoTe$_2$. Although careful measurements reveal the valence band edge to be at $V_{bias}\approx-0.55$~V, the onset of large differential conductance in Fig.~\ref{fig:2}a instead appears at $V_{bias}\approx-0.68$~V. This discrepancy stems from the high variation of the tunneling decay constant, $\kappa$, within the tMoTe$_2$ valence band, which effectively conceals the valence band edge in conventional constant-height d$I$/d$V$ spectroscopy (see Methods for further discussion).

States at the valence band edge originate from the K point of the monolayer MoTe$_2$ Brillouin zone~\cite{Zhang2015, Zhang2020,Zhao2020}, and therefore have a large decay constant in tunneling experiments owing to a large crystal momentum mismatch between the K point and the $s$-wave of an ideal STM tip. An additional contribution to their large decay constant comes from the phase mismatch between the $s$-wave of the tip and the d$_{xy}$+d$_{x^2-y^2}$ orbital character of the K-point states in MoTe$_2$~\cite{monolayer_Liu2013}, as sketched in Fig.~\ref{fig:2}b. In a twisted MoTe$_2$ bilayer, the electronic bands of each monolayer hybridize and are folded into a smaller \moire Brillouin zone (as depicted by the purple hexagon in Fig.~\ref{fig:2}b). As a consequence, the highest energy \moire valence bands share a similarly large decay constant. Fig.~\ref{fig:2}c shows a DFT calculation of the resulting band structure for tMoTe$_2$ with $\theta=2.88^{\circ}$, with bands colored in blue (red) arising from the K ($\Gamma$) point of the monolayer MoTe$_2$ Brillouin zone. The first few valence bands originating from the K point are relatively flat and isolated, and carry non-zero valley Chern numbers, $C_{K\uparrow}=-C_{K'\downarrow}$, where K$\uparrow$ and K'$\downarrow$ represent the locked valley and spin degrees of freedom. States originating from the $\Gamma$ point are well separated from these K-point bands, first emerging at an energy of $\approx150$~meV below the edge of the valence band.

In order to acquire d$I$/d$V$ spectra of the moir\'e flat band states with adequate resolution, we perform a modified version of the conventional constant-height spectroscopy. We first stabilize the tip at a fixed height above the sample, then turn off the feedback loop and move the tip towards the sample by a fixed distance (typically between $0.2-1$~nm), and finally sweep $V_{bias}$ while measuring d$I$/d$V$. We refer to this technique as ``reduced-height spectroscopy'' (see Methods for additional details). The blue curve in Fig.~\ref{fig:2}d shows a representative spectrum acquired in this way. The bump at $V_{bias}\approx-0.59$~V corresponds to the first few flat bands of tMoTe$_2$, all smeared into a single feature. The red curve in Fig.~\ref{fig:2}d shows a conventional constant-height d$I$/d$V$ spectrum acquired from the same area of the sample. In this measurement, d$I$/d$V$ is almost immeasurably small over the range of $V_{bias}$ in which we see the K-band states with reduced-height spectroscopy. Instead, we see a large d$I$/d$V$ peak at $V_{bias}\approx-0.69$~V that arises from bands which originate from the $\Gamma$ point of the monolayer MoTe$_2$ Brillouin zone. These states have small crystal momentum and primarily out-of-plane $d_{z^2}$ character~\cite{Zhao2020}, and thus result in a large d$I$/d$V$ signal compared with K-point states. 

We verify the origin of these features by measuring $\kappa$ as a function of bias (see Fig.~\ref{fig:2}e and Methods). We observe a dip in $\kappa$ coincident with the large d$I$/d$V$ peak seen in the red curve in Fig.~\ref{fig:2}d. This dip is consistent with the prediction that these states primarily arise from bands originating from the $\Gamma$ point. Additionally, the increased decay constant for $V_{bias}\gtrsim-0.68$~V confirms the K-point origin of the valence band edge. The offset between K- and $\Gamma$-point states is consistent with the calculation, in which bands from $\Gamma$ are first found at an energy of $\approx150$~meV below the top of the moir\'e valence band. 

%%%%%%%%%%%%%%%%%%%%%%%%%%%%%%%%%%%%%%%%%%%%%%%%%%%%%%%%%%%%%%%%%%%%%
\begin{figure*}[]
\includegraphics[width=\textwidth]{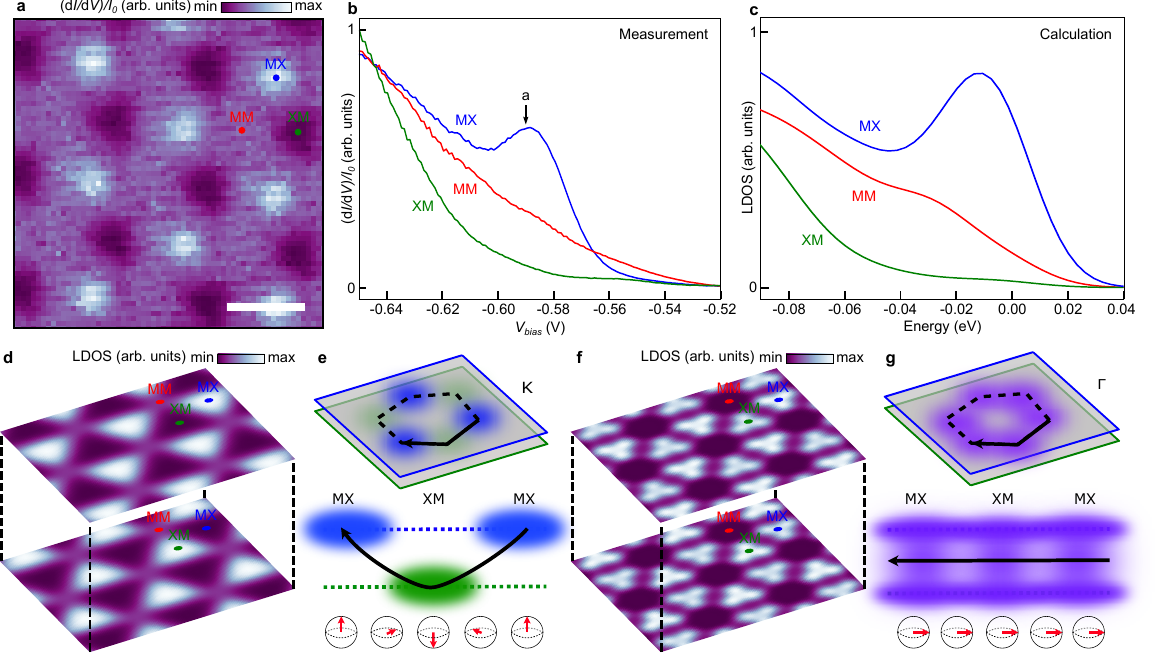} 
\caption{\textbf{Localization of K-point states and connection to band topology.} 
\textbf{a}, d$I$/d$V$ map, normalized by the initial current, $I_0$, acquired using reduced-height spectroscopy with $V_{bias} = -0.59$~V over the same region as in Fig.~\ref{fig:3}c.
\textbf{b}, Reduced-height d$I$/d$V$ spectra normalized by $I_0$ averaged over three different high symmetry regions 
(3x3 pixels, 0.88~nm$^2$). Data in \textbf{a-b} was assembled from grid spectroscopy with initial tunneling setpoint $(I_t, V_{bias}) = (50~\text{pA},-1.0~\text{V})$. The tip was moved towards the sample by a fixed distance of 0.2~nm before the bias sweep.
\textbf{c}, Calculation of the LDOS at MM, MX, and XM for the top layer of tMoTe$_2$ with $\theta=2.88^{\circ}$ over a comparable energy range as the measurement shown in \textbf{b}. The value of energy $E=0$ corresponds to the top of the first moir\'e valence band. A Gaussian smearing is applied to better reflect the experimental measurements.
\textbf{d}, Calculated spatially resolved LDOS at $E=-10~\text{meV}$ for both the top and bottom MoTe$_2$ layers. These states originate from the K points of the monolayer MoTe$_2$ Brillouin zone. The calculation is performed without any external electric field induced by the STM tip (see Extended Data Fig.~\ref{fig:D_dependence} for a similar calculation with this effect included).
\textbf{e}, Schematic representation of the layer-pseudospin winding in the K bands of tMoTe$_2$ along a closed path encircling the moir\'e unit cell. Blue and green shaded regions correspond to wavefunctions polarized to the top and bottom layers, respectively. The red arrow in the Bloch sphere indicates the orientation of the layer pseudospin. 
\textbf{f}, Calculated spatially resolved LDOS at $E=-170~\text{meV}$, corresponding to states arising from the $\Gamma$ point. 
\textbf{g}, Schematic representation of interlayer hybridization of wavefunctions in the $\Gamma$ bands of tMoTe$_2$, and the corresponding Bloch sphere.} 
\label{fig:4}
\end{figure*}
%%%%%%%%%%%%%%%%%%%%%%%%%%%%%%%%%%%%%%%%%%%%%%%%%%%%%%%%%%%%%%%%%%%%%

\medskip\noindent\textbf{Spatial localization of valence band wavefunctions}

Fig.~\ref{fig:3}a shows four representative d$I$/d$V$ spectra taken at different positions within the moir\'e unit cell. These spectra are acquired with standard constant-height spectroscopy, such that observed features originate primarily from $\Gamma$-point moir\'e bands. The green and blue curves are taken at the XM and MX stacking sites, the red at MM, and the pink at the center of a solitonic domain wall (DW) connecting MM sites. We see that these d$I$/d$V$ spectra exhibit considerable spatial variation within the moir\'e unit cell. Fig.~\ref{fig:3}b shows a line-cut of d$I$/d$V$ spectra obtained along the dashed black line in the inset of Fig.~\ref{fig:3}a. From this map, we see a large d$I$/d$V$ peak that is at its lowest energy at the MX and XM sites, and smoothly drifts to its largest energy at the MM sites. 

The spatial variation of the $\Gamma$-point states can be best seen in maps of d$I$/d$V$ acquired at fixed $V_{bias}$ (Fig.~\ref{fig:3}c), which provide a direct visualization of the wavefunction distribution within the moir\'e unit cell and its energy dependence. Consistent with the spectra in Figs.~\ref{fig:3}a-b, we see that d$I$/d$V$ at $V_{bias}=-0.90$V is largest at the MM sites, then shifts to the domain walls as the bias is lowered to $-0.78$~V, further develops into a more complex geometry at $-0.72$~V that is not localized on any high-symmetry stacking site, and finally localizes onto the MX and XM sites at $-0.69$~V. Collectively, these measurements show that the $\Gamma$-point wavefunctions can reside at any position within the moir\'e unit cell, with the details depending sensitively on energy. The precise localization of these states is highly dependent on the exact combination of twist angle, heterostrain, and interlayer separation (see Supplementary Information). Similar behavior was observed over a wide range of twist angles, shown in Extended Data Fig.~\ref{fig:supp_3p5} for 3.52$^\circ$ and Extended Data Fig.~\ref{fig:SupplLowTwistRegion} for 1.20\degree.

In sharp contrast, we find that the flat-band states originating from the K points only reside at certain positions within the \moire unit cell. Fig.~\ref{fig:4}a shows a representative reduced-height d$I$/d$V$ spectroscopy map acquired with $V_{bias}=-0.59$~V (additional maps are shown in Extended Data Fig.~\ref{fig:variedZfull} and Supplementary Video 2). The map is normalized by the initial current of the spectrum, $I_0$, in order to mitigate artifacts inherent to the reduced-height spectroscopy technique (see Supplementary Information for a discussion of this normalization, and Supplementary Information Fig.~S2 for the raw d$I$/d$V$ spectra). The normalized d$I$/d$V$ is largest at the MX stacking site, and becomes progressively weaker at MM and XM. Fig.~\ref{fig:4}b shows the reduced-height spectroscopy spanning a small range of $V_{bias}$ at these three stacking sites. We see that the d$I$/d$V$ peak previously identified in Fig.~\ref{fig:2}d exists only on the MX stacking site, and that d$I$/d$V$ is largest at MX over nearly the entire range of $V_{bias}$ shown (see Supplementary Information for a discussion of the hierarchy of (d$I/$d$V$)/$I_0$ at $V_{bias}\approx-0.55$~V). A replication of this behavior at $\theta=3.48^{\circ}$ is shown in Extended Data Fig.~\ref{fig:variedZ_additional}, and its evolution in the small-$\theta$ limit is shown in Extended Data Fig.~\ref{fig:SupplLowTwist_K}.

\medskip\noindent\textbf{Layer-pseudospin textures and band topology}

The particular spatial localization of the flat bands we observe is well captured by our machine-learning assisted large-scale DFT calculations. Fig.~\ref{fig:4}c shows the calculated LDOS at MX, XM, and MM in the top MoTe$_2$ layer over the same range of energy as in Fig.~\ref{fig:4}b. The calculation is in excellent qualitative agreement with our experimental observations, in particular showing the largest LDOS at MX over the entire range of energy shown. Fig.~\ref{fig:4}d further shows a spatial map of the calculated LDOS for both the top and bottom layers of the twisted MoTe$_2$ (corresponding to an energy of $-10$~meV in Fig.~\ref{fig:4}c). Since tunneling from the STM tip is primarily sensitive to the top layer of tMoTe$_2$, we directly compare our (d$I/$d$V$)/$I_0$ map in Fig.~\ref{fig:4}a with the calculation for the top MoTe$_2$ layer shown in Fig.~\ref{fig:4}d. The calculated LDOS is largest around the MX sites in the top MoTe$_2$ layer, consistent with our experimental observation of largest d$I$/d$V$ at MX.

The layer-dependent localization of the flat-band wavefunctions to different high-symmetry stacking sites in the moir\'e unit cell is the key feature responsible for generating nontrivial band topology. The local layer polarization can be mapped onto a pseudospin Bloch sphere, in which the top (bottom) of the sphere corresponds to a wavefunction polarized to the top (bottom) MoTe$_2$ layer. Under this mapping, the MX site is the north pole and the XM site the south pole. As the electron hops inside the moir\'e unit cell, the trajectory of the layer pseudospin will cover the entire Bloch sphere, giving rise to winding number of 1 (illustrated schematically in Fig.~\ref{fig:4}e)~\cite{hasan_colloquium_2010}. This skyrmion lattice generates a pseudo-magnetic field responsible for the anomalous Hall effect seen in transport experiments~\cite{yu_giant_2020}, analogous to the topological Hall effect driven by skyrmions of electron spin found in certain magnetic materials~\cite{nagaosa_topological_2013}.

The layer-polarizion of the flat-band wavefunctions contrasts that of the $\Gamma$-point states, which can appear at any position within the moir\'e unit cell (Fig.~\ref{fig:3}c). This is anticipated theoretically, since these $\Gamma$-point states are strongly interlayer-hybridized and are thus unable to polarize to a single MoTe$_2$ sheet. Fig.~\ref{fig:4}f shows a calculation of the LDOS at an energy of $170$~meV below the valence band edge, comparable to the d$I$/d$V$ map acquired at $V_{bias}=-0.69$~V in Fig.~\ref{fig:3}c. The experiment and theory agree very well, in that both show wavefunction localization nearly equivalently onto the MX and XM sites in both layers (see Supplementary Information for a discussion of weak $C_{2x}$ symmetry-breaking terms). For these interlayer-hybridized bonding states, the layer pseudospin is pinned to a narrow region on the equator of the Bloch sphere and the resulting texture is topologically trivial (illustrated schematically in Fig.~\ref{fig:4}g, with the direction along the equator chosen arbitrarily).

A potential complication in interpreting our measurements lies in understanding the effect of a local electric field under the STM tip, arising from the large $V_{bias}$ and an intrinsic work function mismatch between the tip and the sample. We have performed additional DFT calculations similar to those described above, but also including a realistic tip-induced external electric field (see Methods and Extended Data Fig.~\ref{fig:D_dependence}). We find that the states originating from the $\Gamma$ point have an almost negligible dependence on the external field owing to their interlayer hybridization. In contrast, the relative weight of the K-point states shifts towards one of the two MoTe$_2$ layers, depending on the sign of the external field. Nevertheless, the geometry of the LDOS localization within the moir\'e unit cell remains unchanged. Thus, a tip-induced field does not meaningfully impact the primary conclusions of our analysis.

\medskip\noindent\textbf{Discussion and outlook}

Our STM/S study of tMoTe$_2$ provides the first microscopic visualization of this unique topological material. Accurately predicting the Chern numbers and quantum geometry of the flat valence bands has been challenging owing to their extreme sensitivity to the modeling parameters~\cite{Wu2018_topological,Zhang_Polarization_2023, Wang2024,jia_moire_2023,ahn_first_2024}. Our detailed energy-dependent LDOS maps and atomically-resolved topographs now provide strong experimental constraints for theory. In particular, our observation of MX-localized wavefunctions in the top MoTe$_2$ layer is in excellent agreement with our modeling. Our theory further reveals that the in-plane atomic lattice relaxations within the moir\'e lattice are instrumental in controlling the topological properties of the material. In calculations with in-plane lattice relaxations artificially excluded, the north and south poles of the pseudospin lattice flip positions between MX and XM and also change the Chern numbers of the flat bands (Extended Data Fig.~\ref{fig:no_relaxations}). This new understanding unveils the intimate connection between the precise atomic lattice structure of tMoTe$_2$ and its quantum geometric properties.

Looking forward, the structure of the LDOS within the moir\'e unit cell and its evolution with energy can now be used to further refine theories beyond simply predicting the correct Chern number. This will help to provide sharper constraints on the FQAH effect in tMoTe$_2$, likely resulting in a better understanding of experimentally observed states and aiding predictions for entirely new phases. Our results also provide insights into the atomic-scale electric polarization structure of tMoTe$_2$ arising from the combination of microscopic ferroelectric and piezoelectric contributions, informing ongoing studies of the moir\'e ferroelectricity found in twisted TMD bilayers~\cite{Zhang_Polarization_2023,McGilly2020,Wang2022,Molino2023,Zhang2023_plasmons} and opening new pathways towards engineering the FQAH states with strain~\cite{Liu2024_strain}. Lastly, our foundational atomic-scale understanding of the system enables future STM/S experiments probing the intertwined correlated and topological states that emerge upon hole doping. Progress in this direction will require the development of new device geometries with Ohmic contacts to the tMoTe$_2$, obviating the need for a graphene substrate and enabling gating to the integer and fractional Chern insulator states.

\section*{Methods}

\textbf{Device fabrication.} 
Our vdW hetereostructures are fabricated using a standard dry transfer technique with a poly(bisphenol A carbonate) (PC) film on a polydimethylsiloxane (PDMS) puck. Flakes of graphite, hBN, and 2H-MoTe$_2$ are exfoliated onto a SiO$_2$/Si wafer and are then picked up using the PC stamp in the following order: hBN, monolayer graphene, tMoTe$_2$, graphite. The graphite covers only a portion of the graphene, and extends beyond the hBN to eventually establish electrical connection to the STM wiring. To assemble the tMoTe$_2$, we first identify two adjacent monolayer MoTe$_2$ flakes (which likely have the same orientation of their crystal axes), or cut a single monolayer MoTe$_2$ flake in half using a sharp metal tip affixed to a micromanipulator. We then pick up the two pieces of MoTe$_2$ sequentially, rotating the sample stage by the desired twist angle after picking up the first piece. After the entire vdW heterostructure is assembled, the PC film is peeled off the PDMS puck by hand, and carefully placed upside-down onto a different PDMS puck. At this point, the tMoTe$_2$ is in direct contact with the PDMS, and the entire vdW stack is covered by PC. The PC film is then dissolved using N-Methyl-2-pyrrolidone (NMP) and subsequently washed with isopropyl alcohol (IPA), leaving just the vdW stack resting on PDMS. 

In parallel, a SiO$_2$/Si substrate is pre-patterned with evaporated metal electrodes. These electrodes are fabricated using standard optical lithography and electron-beam metal evaporation (5~nm of Cr followed by 50~nm of Au) techniques. The wafer is attached to the STM sample plate using silver epoxy, and wires are hand-pasted between the evaporated Cr/Au electrodes and the electrical contacts on the STM sample plate. Next, the vdW stack is deposited directly onto the Cr/Au electrodes by pressing down the PDMS puck onto the substrate. The graphite contacts are oriented such that they land on the pre-patterned Cr/Au electrodes, taking care not to unintentionally short electrodes together. The vdW stack delaminates easily from the PDMS, and as a result it is flipped upside down from its original assembly such that the tMoTe$_2$ is furthest from the substrate and remains exposed for STM measurements. Lastly, any residual PDMS contamination is removed from the surface of the tMoTe$_2$ by densely scanning with an atomic force microscope (AFM) tip operated in contact mode. Our technique is illustrated in Extended Data Fig.~\ref{fig:image_fab}a, and is a modified version of that first introduced in Ref.~\cite{Choi2019}. Extended Data Fig.~\ref{fig:image_fab}b shows optical images of the final devices used in this study. The MoTe$_2$ flakes for device D1 are from homegrown crystals grown using Te flux, those for device D2, D3, and D4 are from commercial crystals from HQ Graphene. The data for the $\theta=2.75^{\circ}$ region is from device D1, and all other data is from devices D2, D3, and D4.

Because MoTe$_2$ degrades when exposed to air, all vdW device assembly steps involving MoTe$_2$ are performed in an argon environment in a glovebox ($<$0.1 ppm O$_2$ and H$_2$O). The completed device is transferred directly from the glovebox into the STM load lock chamber using a vacuum transfer suitcase. This process results in clean MoTe$_2$ surface that is almost entirely free of residue. Fig.~\ref{fig:image_fab}c shows a representative STM topograph of a 100~nm by 100~nm region of a tMoTe$_2$ sample, showing that atomic defects in the MoTe$_2$ are the primary form of disorder in our samples (rather than sample degradation).

\textbf{Effects of the graphene substrate.} We use a graphene substrate in our devices owing to challenges in otherwise forming Ohmic contact to tMoTe$_2$ at cryogenic temperatures. We find that monolayer graphene efficiently collects electrons tunneled into the tMoTe$_2$. Because of the large mismatch in the atomic lattice constants of graphene and MoTe$_2$, states at the K points of each material are separated by a large crystal momentum and do not strongly hybridize. We estimate the size of the Fermi pocket around the K point in tMoTe$_2$ from the calculated dispersion of untwisted bilayer MoTe$_2$. Similarly, we estimate the size of the Fermi pocket in graphene by assuming a linear dispersion, $E=\hbar v_{F} k$, with $v_F=1.1\times 10^6$~m/s~\cite{wallace_band_1947}. We find that at energies up to 1~eV below the valence band edge of tMoTe$_2$, and with no interlayer twist, the Fermi pockets of MoTe$_2$ and graphene are separated by at least 2~nm$^{-1}$ and thus do not hybridize. Twisting the two materials further separates the low-energy states. Therefore, we expect that the primary effects of the graphene substrate compared to a standard dielectric (such as hBN) are: (i) creating an isopotential surface for the bottom MoTe$_2$ layer, and (ii) weakly modifying the precise form of the tMoTe$_2$ lattice relaxations (see Supplementary Information for additional discussion). Lastly, at a practical level, the graphene substrate also precludes gating experiments since it partially screens the electric field effect from the silicon back gate.

\textbf{STM/STS measurements.} 
Our measurements are performed using a Scienta Omicron Polar STM in an ultrahigh vacuum cryogenic environment with base temperature of $T=4.8$~K. The sample plate has ten isolated electrical contacts, enabling the use of either scanning gate microscopy or capacitive searching~\cite{andrei_capa} techniques to navigate to the small tMoTe$_2$ sample. In our experimental setup, the tMoTe$_2$ sample is grounded and a voltage bias is applied to the tip. To facilitate the interpretation of our d$I$/d$V$ spectra, we plot the data in terms of sample bias, which is related to the tip bias as follows: $V_{bias}=-V_{tip}$. The STM tip is made by electrochemically etching a tungsten wire. The end of the STM tip is shaped on the evaporated gold electrode nearby the tMoTe$_2$ by pulsing and poking into the gold until the STM scanning is reliable and the $I(V)$ curve exhibits a perfectly linear slope. 

The d$I$/d$V$ spectra are acquired using a phase-sensitive detection with a $V_{mod} = 5$~mV peak-to-peak AC voltage modulation at a frequency of either 317~Hz or 419~Hz. The first harmonic of the demodulated signal is proportional to d$I$/d$V$ and thus to the local density of states. 
The energy resolution for spectroscopy is $\Delta E = \sqrt{(3.5 k_B T )^2 + (eV_{mod})^2 }\approx5$~meV where the first term corresponds to the thermal broadening and the second accounts for the instrumental broadening from the lock-in modulation. Note that this broadening is underestimated because the transmission of tunneling quasiparticles grows exponentially with bias, so that at larger bias there is a high number of tunneling channels that make d$I$/d$V$ sensitive to the LDOS over a broader range of energy~\cite{lang_spectroscopy_1986}.

We use either conventional constant-height spectroscopy or a modified ``reduced-height'' spectroscopy technique to probe electronic states, depending on the relative magnitude of their decay constant. In particular, larger $\kappa$ is observed for states originating from $K$-point bands due both to the in-plane orbital character of these states~\cite{monolayer_Liu2013} and to their higher crystal momentum $k_\parallel$, following equation $\kappa = \sqrt{\frac{2m(\Phi_0-E-V/2)}{\hbar^2} + k_\parallel^2}$ \cite{TersoffHamann,TosattiChen}. Here, $\Phi_0$ is the average work function of the tip and the sample, $V$ is the bias and $E$ the energy of the tunneling electrons involved.

To perform constant-height spectroscopy, the tip is first stabilized at a given current and bias setpoint, $(I_t,V_{bias})$. Then the feedback loop is turned off and $V_{bias}$ is swept across the range of interest while measuring d$I/$d$V$. Such measurements can be acquired over a spatially resolved grid of points, generating a series of LDOS maps with high energy resolution, high tip stability, and minimal tip-induced doping (since the tip is kept relatively far the sample at all times). These grid spectroscopy measurements are analyzed using STMaPy (Scanning Tunneling Microscopy analysis in Python), a data analysis tool written in Python~\cite{stmapy}. We use this technique mainly to study bands originating from the $\Gamma$ point.

When performing reduced-height spectroscopy, the spectra are acquired in the same way, however the tip is moved closer to the sample by a fixed distance, $\Delta z$, before $V_{bias}$ is swept. This allows for stabilization far away from the sample, avoiding artifacts stemming from position-dependent height variations and tip-sample interactions that arise when the tip is stabilized closer to the sample (see Supplementary Information). The pre-measurement height reduction serves to enhance the d$I$/d$V$ signal from states with larger $\kappa$ originating from the K points. Because small fluctuations in the stabilization condition can be exponentially amplified at a decreased tip-sample separation, we normalize individual spectra by the initial current, $I_0$, to make comparisons across spectra more consistent (see Supplementary Information for a discussion on the normalization).

\textbf{Decay constant measurements.} 
Decay constant measurements are performed by fitting measurements of $I(z)$ at different $V_{bias}$. For each bias voltage, the tip is first stabilized at a current setpoint of 10~pA and then gradually retracted by a distance of 1~nm while the tunneling current is measured. $I(z)$ curves are fit using $I(z) \propto e^{-2\kappa z}$, where $z$ is the tip-sample separation and $\kappa$ is the decay constant.

\textbf{Determination of intrinsic heterostrain in the experiment.}
All topography images acquired by STM are analysed using Gwyddion and Inkscape. Following the procedure described in Refs.~\cite{Artaud2016, huder_electronic_2018}, the relative strain between the MoTe$_2$ layers is determined in the most general case by solving the following equation relating the top layer atomic lattice, $\bm{a_{t_i}},i=1,2$, and the bottom layer atomic lattice, $\bm{a_{b_i}},i=1,2$, as determined from atomically resolved images: 

\begin{align}
	\begin{pmatrix} \bm{a_{t_1}} \\ \bm{a_{t_2}} \end{pmatrix}
 =&
 \begin{pmatrix} cos\theta_2 & sin\theta_2 \\ -sin\theta_2 & cos\theta_2 \end{pmatrix} \begin{pmatrix} 1+\varepsilon_{uni} & 0 \\0 & 1 \end{pmatrix} \nonumber\\ 
 &\begin{pmatrix} cos\theta_1 & sin\theta_1 \\ -sin\theta_1 & cos\theta_1 \end{pmatrix} \begin{pmatrix} 1+\varepsilon_{bi} & 0 \\0 & 1+\varepsilon_{bi} \end{pmatrix}\begin{pmatrix} \bm{a_{b_1}} \\ \bm{a_{b_2}} \end{pmatrix}.
	\label{eq:Park Madden2}
\end{align}

In this scheme, starting with the bottom lattice which is assumed to be undeformed, a biaxial deformation quantified by $\varepsilon_{bi}$ is first applied, then the lattice is rotated by $\theta_1$, after which a uniaxial deformation quantified by $\varepsilon_{uni}$ is applied in the original $\bm{a_{b_1}}$ direction, and finally the lattice is further rotated by $\theta_2$, yielding the top lattice. The twist angle is thus $\theta=\theta_1+\theta_2$, and the uniaxial strain is applied in the direction $\theta_1$ with respect to the \moire wavevector.

For STM images without atomic resolution, we follow reference \cite{Kerelski2019} in assuming Poisson behaviour of the lattice and minimize the following equation using the moir\'e lengths as input parameters:

\begin{align}
	\begin{pmatrix} \bm{a_{t_1}} \\ \bm{a_{t_2}} \end{pmatrix}  =&  \begin{pmatrix} cos\theta_s & -sin\theta_s \\ sin\theta_s & cos\theta_s \end{pmatrix}
    \begin{pmatrix} 1+ \varepsilon & 0 \\0 & 1-\delta \varepsilon \end{pmatrix} \nonumber\\& 
    \begin{pmatrix} cos\theta_s & sin\theta_s \\ -sin\theta_s & cos\theta_s \end{pmatrix}
    \begin{pmatrix} cos\theta & sin\theta \\ -sin\theta & cos\theta \end{pmatrix} \begin{pmatrix} \bm{a_{b_1}} \\ \bm{a_{b_2}} \end{pmatrix} ,
	\label{eq:4transfokerel}
\end{align}
with the MoTe$_2$ Poisson ratio estimated to be $\delta = 0.37$~\cite{Duerloo2012}, and $\varepsilon$ the anisotropic heterostrain oriented in the direction $\theta_s$ with respect to the \moire lattice vectors. This method is used for the regions from Fig.~\ref{fig:1}e, Extended Data Fig.~\ref{fig:SupplLowTwistRegion}, and Extended Data Fig.~\ref{fig:SupplLowTwist_K}. In this case, we are sensitive to piezo calibration artifacts and overlook biaxial heterostrain. The latter has been consistently found to be non-negligible in other regions of the sample, and is therefore the main source of error in the determination of the twist angle, which is likely overestimated here by a fraction of a degree.

\textbf{DFT calculations.}
We use neural network (NN) potentials to perform lattice relaxations of moir\'{e} superlattices. The NN potentials are parameterized using the deep potential molecular dynamics (DPMD) method~\cite{zhang2018deep,wang2018deepmd}. To generate training datasets, we perform 5000-step \textit{ab initio} molecular dynamics (AIMD) simulations for $\theta=6^\circ$ tMoTe$_2$ at $T=500$~K using the \textsc{vasp} package~\cite{kresse1996efficiency}, with van der Waals corrections considered within the D2 formalism~\cite{grimme2006semiempirical}. More details on parameterizing NN potentials can be found in Ref.~\cite{Zhang_Polarization_2023}. Subsequently, the trained NN potentials are employed to relax the moir\'{e} superlattice using the \textsc{lammps} package~\cite{thompson2022lammps} until the maximum atomic force is less than 10$^{-4}$~eV/\AA.

The moir\'{e} band structures and LDOS are calculated using the \textsc{siesta} package~\cite{soler2002siesta}, with the inclusion of spin-orbit coupling. We employ the optimized norm-conserving Vanderbilt (ONCV) pseudopotential~\cite{hamann2013optimized}, the Perdew-Burke-Ernzerhof (PBE) functional~\cite{perdew1996generalized}, and a double-zeta plus polarization basis. To sample the moir\'{e} Brillouin zone for LDOS calculations, a $k$-grid of $3\times3\times1$ is utilized, where $k$ is the wavevector. The real-space distribution of the wavefunction is approximated by the weight of projected wavefunction onto the Mo and Te atomic orbitals. Gaussian smearing with a width of 33~meV is applied to smooth the LDOS curves for comparison to experimental data. For the calculated LDOS curves corresponding to specific stacking sites, we average LDOS values within a circle of radius 10~\AA{} for MX, XM, and MM.

\textbf{Effects of a tip-induced electric field on the K- and $\Gamma$-point wavefunctions.}
The large work function mismatch between gold (which coats the end of the tungsten STM tip) and MoTe$_2$~\cite{workfunction-Aftab,workfunction-Mleczko} creates a substantial electric field between the tip and grounded graphene. This, in conjunction with the applied tip bias, can create an electric field under the tip that locally dopes the tMoTe$_2$ and deforms the \moire bands. Previous studies of tMoTe$_2$ have shown that an electric field can modify the nature of the correlated states~\cite{Cai2023,Park2023,Zeng2023,Fan2023}. Extended Data Figs.~\ref{fig:D_dependence}a-b show zoom-ins of the \moire band structure near the band edge for both $D=0$ and $300~\mathrm{mV/nm}$. Although the $D=0$ band structure features several Chern bands, the presence of a finite electric displacement field, $D$, causes band inversions and splittings that result in topologically trivial bands. 

To investigate the dependence of the LDOS structure on $D$, we calculate the LDOS at different high symmetry points in both MoTe$_2$ layers for both zero and finite $D$ as shown in Extended Data Fig.~\ref{fig:D_dependence}c-d (solid curves correspond to the top layer, and dashed to the bottom). The main consequence of a finite $D$ is an energy splitting between the peak at MX in the top layer and the peak at XM in the bottom layer, indicating a slight layer polarization. The calculated LDOS spatial maps in Extended Data Fig.~\ref{fig:D_dependence}e further illustrate that $D$ induces some degree of layer polarization. Despite this, we see that the system retains the overall geometry of LDOS localization at MX in the top layer, as seen at $D=0$ in Fig.~\ref{fig:4}d. Although finite $D$ can modify the Chern number of the bands to zero, the LDOS geometry we observe is nevertheless indicative of flat bands that can have a finite Chern number at $D=0$. In contrast, spatial maps of the LDOS at the $\Gamma$ point at finite $D$, shown in Extended Data Fig.~\ref{fig:D_dependence}f, are nearly indistinguishable from those at $D=0$ due to strong interlayer hybridization. In summary, the LDOS localization we see provides detailed information to compare with band structure calculations predicting the topology of the bands, though the experimental data does not uniquely indicate the Chern number. 

\section*{Acknowledgements}
We thank D. Waters and B. LeRoy for valuable technical discussions. Experimental and theoretical research on the topological properties of twisted molybdenum ditelluride is supported as part of Programmable Quantum Materials, an Energy Frontier Research Center funded by the U.S. Department of Energy (DOE), Office of Science, Basic Energy Sciences (BES), under award DE-SC0019443. The development of twisted molybdenum ditelluride samples and their basic STM characterization is supported by the U.S. Department of Energy, Office of Science, Office of Basic Energy Sciences under Award Number DE-SC0023062. X.X. and M.Y. acknowledge support from the State of Washington-funded Clean Energy Institute. E.T. was supported by grant no. NSF GRFP DGE-2140004. This work made use of shared fabrication facilities provided by NSF MRSEC 1719797. Machine learning and first-principles calculations are in-part supported by the discovering AI@UW Initiative and AI-Core of the Molecular Engineering Materials Center at the University of Washington (DMR-2308979). This work was facilitated through the use of advanced computational, storage, and networking infrastructure provided by the Hyak supercomputer system and funded by DMR-2308979. K.W. and T.T. acknowledge support from the Elemental Strategy Initiative conducted by the MEXT, Japan (grant no. JPMXP0112101001) and JSPS KAKENHI (grant nos. 19H05790, 20H00354 and 21H05233).

\textbf{Author contributions.} E.T., K.T.C., and F.M. fabricated the devices and performed the measurements.  X.-W.Z. performed the DFT calculations under the supervision of T.C. and D.X. C.H., Y.Z., J.Y., and J.-H.C. grew some of the MoTe$_2$ crystals used in this study. K.W. and T.T. grew the BN crystals. H.P., J.C., E.A., and X.X. provided valuable discussion on device fabrication and data interpretation. M.Y. supervised the project. E.T., K.T.C., F.M., and M.Y. wrote the paper with input from all authors.

\section*{Competing interests}
The authors declare no competing interests.

\section*{Additional Information}
Correspondence and requests for materials should be addressed to Ting Cao, Di Xiao or Matthew Yankowitz.

\section*{Data Availability}
Source data are available for this paper. All other data that support the findings of this study are available from the corresponding author upon request.

\bibliographystyle{naturemag}
\bibliography{references}

\clearpage

\renewcommand{\figurename}{Extended Data Fig.}
\renewcommand{\thesubsection}{S\arabic{subsection}}
\setcounter{secnumdepth}{2}
%\renewcommand{\theequation}{S\arabic{equation}}
%\renewcommand{\thetable}{S\arabic{table}}
%\subsubsectionfont{\normalfont\large\itshape\underline}
\setcounter{figure}{0} 
\setcounter{equation}{0}

\onecolumngrid

\section*{Extended Data}

%%%%%%%%%%%%%%%%%%%%%%%%%%%%%%%%%%%%%%%%%%%%%%%%%%%%%%%%%%%%%%%%%%%%%
\begin{figure*}[h]
\includegraphics[width=\textwidth]{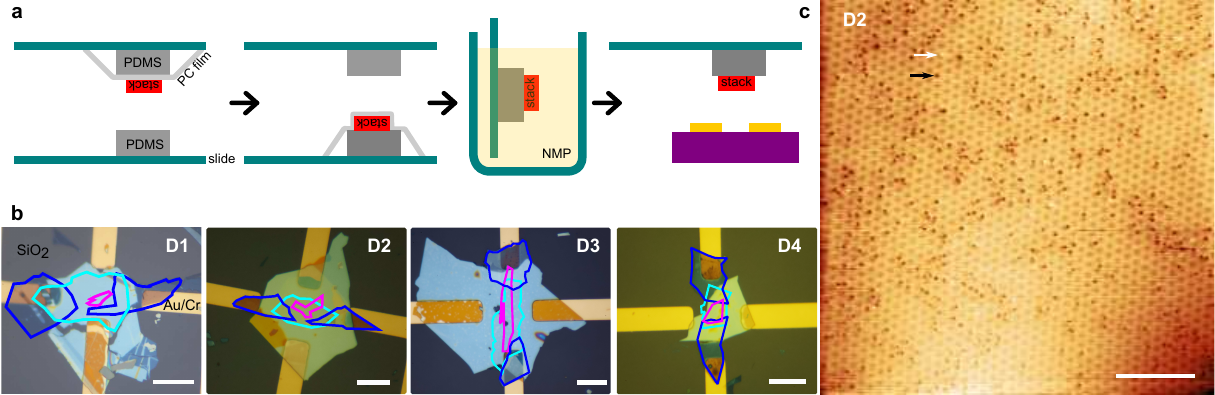}
\caption{\textbf{Device fabrication and large-area STM topograph.} 
\textbf{a}, Schematic illustration of the device fabrication process (see Methods for details).
\textbf{b}, Optical micrographs of the devices used in this study. Flake boundaries are outlined using the following convention: blue - graphite, cyan - monolayer graphene, pink - tMoTe$_2$. Scale bars are 20~$\mu$m. All results in this study were obtained on device D1, except for the following: Fig.~\ref{fig:1}g, Extended Data Figs.~\ref{fig:image_fab}c,~\ref{fig:supp_3p5},~\ref{fig:variedZ_additional}, and Supplementary Information Figs.~S2d-f,~S5 (device D2); Fig.~\ref{fig:1}e, Extended Data Figs.~\ref{fig:SupplLowTwistRegion},~\ref{fig:SupplLowTwist_K} (device D3); Supplementary Information Fig.~S3a-c (device D4).
\textbf{c}, STM topograph of a large area of clean monolayer MoTe$_2$ showing two types of atomic defects~\cite{Edelberg2019}, bright and dark, emphasized by the white and black arrows. A graphene-hBN \moire pattern is also visible in this region. $(I_t,V_{bias})=(100~\text{pA},-1.6~\text{V})$, scale bar is 40~nm.
}
\label{fig:image_fab}
\end{figure*}
%%%%%%%%%%%%%%%%%%%%%%%%%%%%%%%%%%%%%%%%%%%%%%%%%%%%%%%%%%%%%%%%%%%%%

%%%%%%%%%%%%%%%%%%%%%%%%%%%%%%%%%%%%%%%%%%%%%%%%%%%%%%%%%%%%%%%%%%%%%
\begin{figure*}[h]
\includegraphics[width=\textwidth]{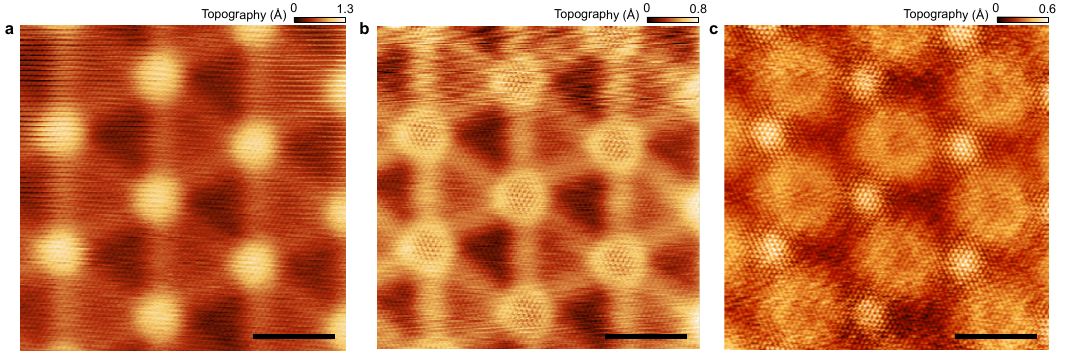}
\caption{\textbf{Dependendence of STM topography on tunneling parameters.} 
STM topographs of the same $\theta=2.75^{\circ}$ region as shown in Fig.~\ref{fig:1}f, but with different tunneling setpoint parameters $(I_t,V_{bias})$: \textbf{a}, $(50~\text{pA}, -1.0~\text{V})$, \textbf{b}, $(110~\text{pA}, -0.9~\text{V})$, and \textbf{c}, $(300~\text{pA}, -0.8~\text{V})$. Scale bars are 5~nm.}
\label{fig:topoD1}
\end{figure*}
%%%%%%%%%%%%%%%%%%%%%%%%%%%%%%%%%%%%%%%%%%%%%%%%%%%%%%%%%%%%%%%%%%%%%

%%%%%%%%%%%%%
%%%%%%%%%%%%%%%%%%%%%%%%%%%%%%%%%%%%%%%%%%%%%%%%%%%%%%%%
\begin{figure*}[h]
\includegraphics[width=\textwidth]{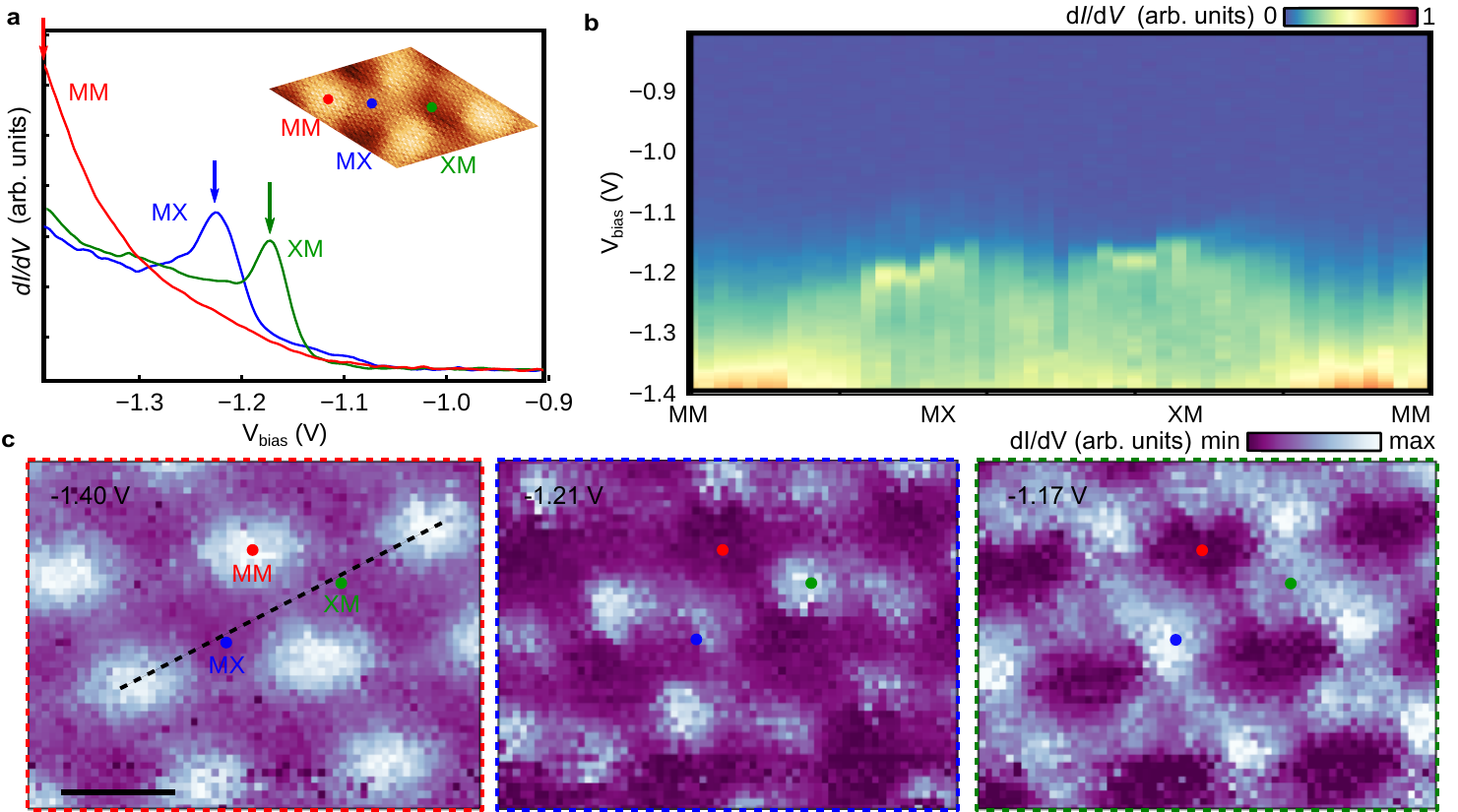} 
\caption{\textbf{Localization of $\Gamma$-point states for tMoTe$_2$ with $\theta=3.52^{\circ}$.}
\textbf{a}, Averaged constant-height d$I$/d$V$ spectra at MM, MX, and XM (4x4 pixels, 1.24~nm$^2$). The inset shows the STM topograph of the region, with $\varepsilon_{uni}=1.90\%$, and $\varepsilon_{bi}=0.24\%$, oriented at $-21^{\circ}$ with respect to the \moire. The tunneling parameters are $(I_t,V_{bias})=(140~\text{pA},-2.5~\text{V})$. 
\textbf{b}, Line cut of constant-height d$I$/d$V$ spectra acquired along the dashed path shown in \textbf{c}.  
\textbf{c}, d$I$/d$V$ maps acquired at different $V_{bias}$, as indicated by the arrows in \textbf{a}. All maps are assembled from a grid spectroscopy measurement with initial tunneling setpoint $(I_t, V_{bias}) = (200~\text{pA},-1.4~\text{V})$. The scale bar is 5~nm.}
\label{fig:supp_3p5}
\end{figure*}
%%%%%%%%%%%%%%%%%%%%%%%%%%%%%%%%%%%%%%%%%%%%%%%%%%%%%%%%%%%%%%%%%%%%%

%%%%%%%%%%%%%%%%%%%%%%%%%%%%%%%%%%%%%%%%%%%%%%%%%%%%%%%%%%%%%%%%%%%%%
\begin{figure*}[h]
\includegraphics[width=\textwidth]{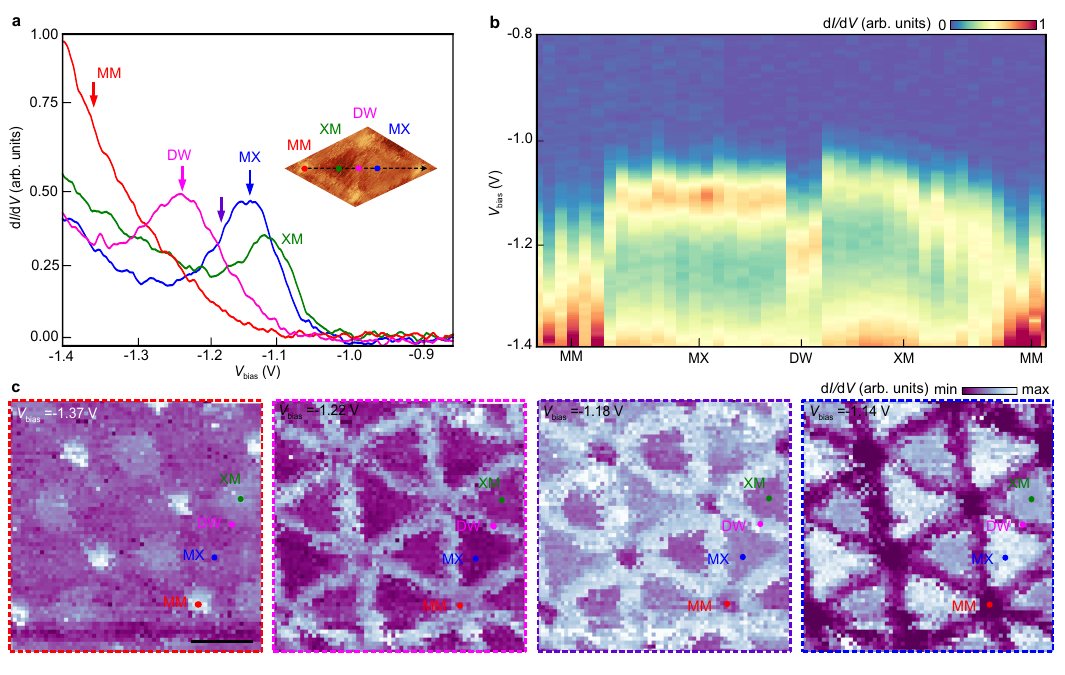} 
\caption{\textbf{Localization of $\Gamma$-point states for tMoTe$_2$ with $\theta=1.20^{\circ}$.} \textbf{a}, Constant-height d$I$/d$V$ spectra at various points within the moir\'e unit cell, acquired at the same region as Fig.~\ref{fig:1}e. The inset shows the STM topograph of the region. 
\textbf{b}, Line cut of constant-height d$I$/d$V$ spectra acquired along the dashed path shown in the inset of \textbf{a}.
\textbf{c}, d$I$/d$V$ maps acquired at different $V_{bias}$, as indicated by the arrows in \textbf{a}. All maps are assembled from a grid spectroscopy measurement with initial tunneling setpoint $(I_t, V_b)=(200~\text{pA},-1.4~\text{V})$. The scale bar is 16~nm.}
\label{fig:SupplLowTwistRegion}
\end{figure*}
%%%%%%%%%%%%%%%%%%%%%%%%%%%%%%%%%%%%%%%%%%%%%%%%%%%%%%%%%%%%%%%%%%%%%

%%%%%%%%%%%%%%%%%%%%%%%%%%%%%%%%%%%%%%%%%%%%%%%%%%%%%%%%%%%%%%%%%%%%%
\begin{figure*}[h]
\includegraphics[width=\textwidth]{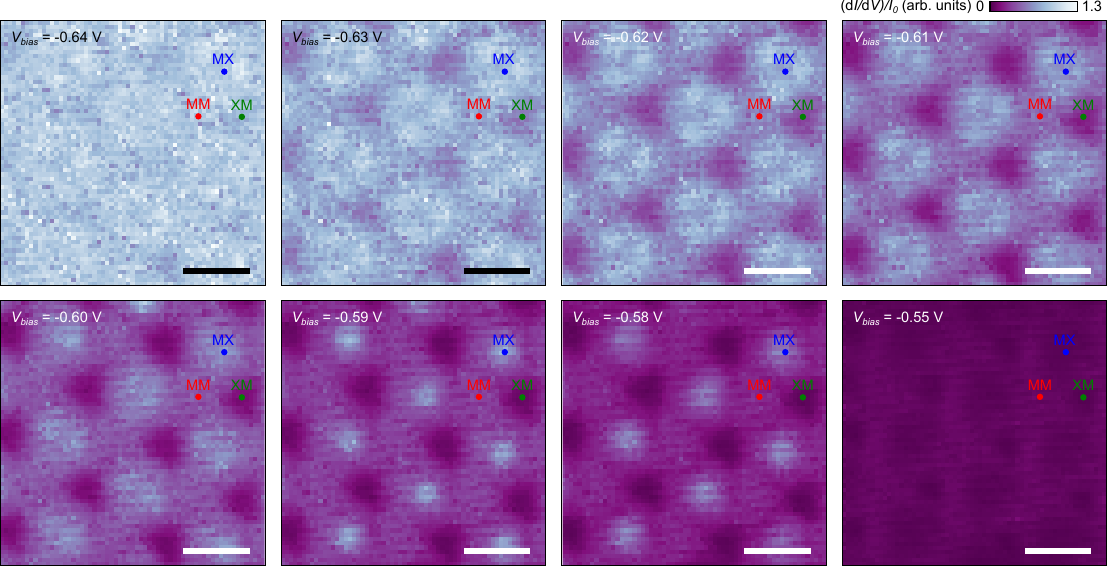}
\caption{\textbf{Additional d$I$/d$V$ maps of K-point states in the $\theta=2.75^\circ$ sample.} Reduced-height d$I$/d$V$ maps in the same region as Fig.~\ref{fig:4}a, shown for additional values of $V_{bias}$ with the same color scale across all maps. Scale bars are 5~nm. See Supplementary Video 2 for an animation of the full evolution with $V_{bias}$, and Supplementary Information for a discussion about the remnant signal at $V_{bias}=-0.55$~V.}
\label{fig:variedZfull}
\end{figure*}
%%%%%%%%%%%%%%%%%%%%%%%%%%%%%%%%%%%%%%%%%%%%%%%%%%%%%%%%%%%%%%%%%%%%%

%%%%%%%%%%%%%%%%%%%%%%%%%%%%%%%%%%%%%%%%%%%%%%%%%%%%%%%%%%%%%%%%%%%%%
\begin{figure*}[h]
\includegraphics[width=\textwidth]{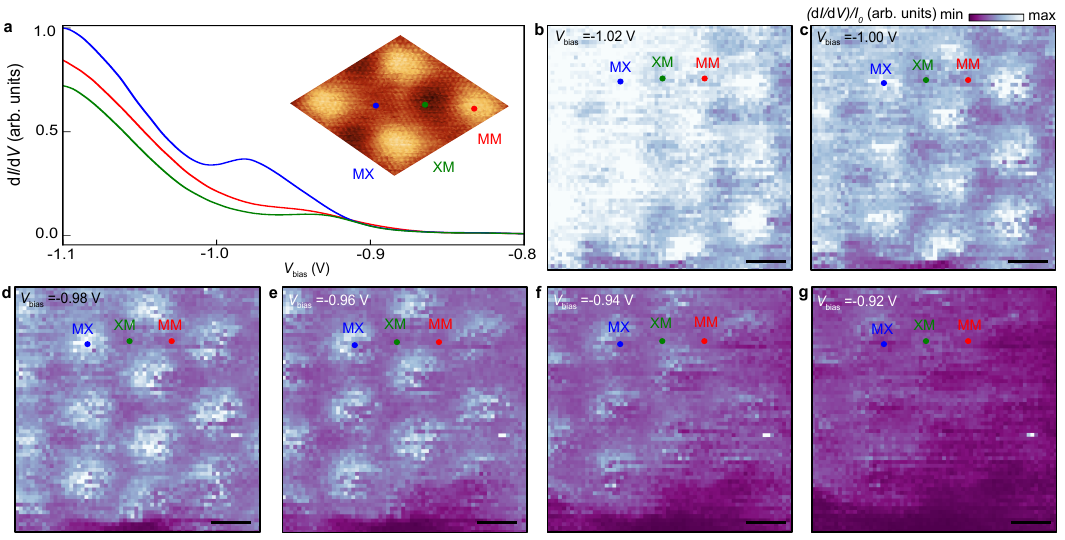}
\caption{\textbf{Localization of K-point states for tMoTe$_2$ with $\theta=3.48^{\circ}$.}
\textbf{a}, Averaged reduced-height d$I$/d$V$ spectra at MM, MX, and XM (3x3 pixels, 0.71~nm$^2$). The inset shows the STM topograph of the region, with $\varepsilon_{uni}=0.74\%$, and $\varepsilon_{bi}=0.00\%$, oriented at $2^{\circ}$ with respect to the \moire. The tunneling parameters are $(I_t,V_{bias})=(50~\text{pA},-2.0~\text{V})$.  
\textbf{b-g}, d$I$/d$V$ maps acquired at different values of $V_{bias}$, as specified in each panel, with the same color scale across all maps. All maps are assembled from a grid spectroscopy measurement with initial tunneling setpoint $(I_t,V_{bias})=(10~\text{pA},-1.2~\text{V})$. The tip was moved towards the sample by a fixed distance of 0.36~nm before the bias sweep. The scale bars are all 3~nm. See Supplementary Video 3 for animation of the full evolution with $V_{bias}$.}
\label{fig:variedZ_additional}
\end{figure*}
%%%%%%%%%%%%%%%%%%%%%%%%%%%%%%%%%%%%%%%%%%%%%%%%%%%%%%%%%%%%%%%%%%%%%

%%%%%%%%%%%%%%%%%%%%%%%%%%%%%%%%%%%%%%%%%%%%%%%%%%%%%%%%%%%%%%%%%%%%%
\begin{figure*}[h]
\includegraphics[width=\textwidth]{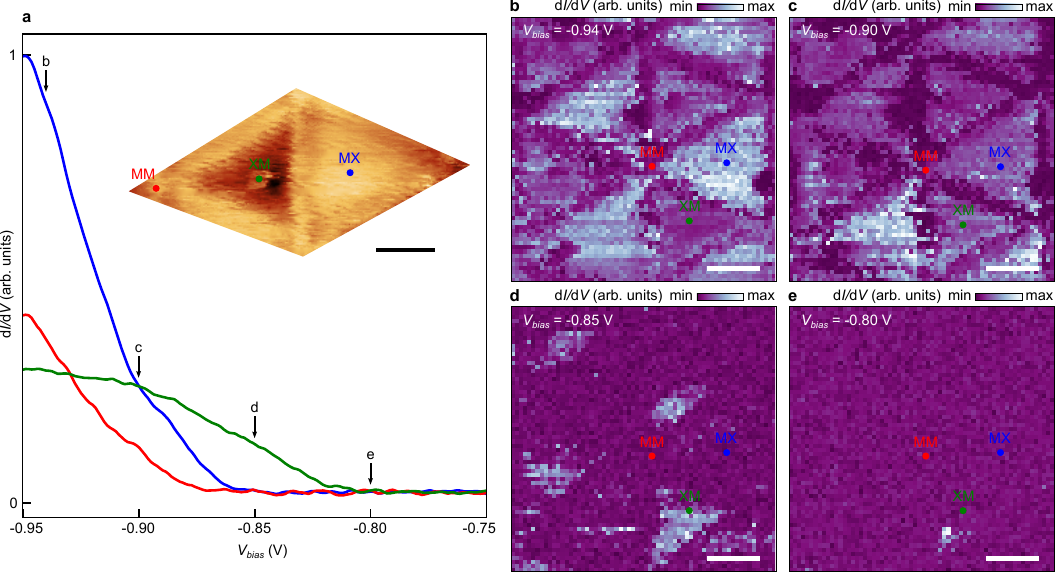}
\caption{\textbf{Localization of K-point states for tMoTe$_2$ with $\theta=0.80^{\circ}$.} \textbf{a}, Averaged reduced-height d$I$/d$V$ spectra at MM (3x3 pixels, 5.49~nm$^2$), MX and XM (4x4 pixels, 9.77~nm$^2$). The inset shows the STM topograph of the region, with $\varepsilon=0.14\%$ oriented at $35^\circ$ with respect to the \moire. 
\textbf{b-e}, d$I$/d$V$ maps acquired at different values of $V_{bias}$, as specified in each panel and by the arrows in \textbf{a}. All maps are assembled from a grid spectroscopy measurement with initial tunneling setpoint $(I_t,V_{bias})=(50~\text{pA},-1.2~\text{V})$. The tip was moved towards the sample by a fixed distance of 0.45~nm before the bias sweep. The scale bars are all 10~nm. See Supplementary Video 4 for animation of the full evolution with $V_{bias}$. These measurements suggest that at small twist angles, the localization of states near the valence band edge switches to XM. However, it is not definitive proof due to complications arising from atomic-scale defects, which become unavoidable at small twist angle due to the very large moir\'e unit cell. Some of these defects can be seen in \textbf{b} and \textbf{e} as dark and bright spots.}
\label{fig:SupplLowTwist_K}
\end{figure*}
%%%%%%%%%%%%%%%%%%%%%%%%%%%%%%%%%%%%%%%%%%%%%%%%%%%%%%%%%%%%%%%%%%%%%

%%%%%%%%%%%%%%%%%%%%%%%%%%%%%%%%%%%%%%%%%%%%%%%%%%%%%%%%%%%%%%%%%%%%%
\begin{figure*}
\includegraphics[width=\textwidth]{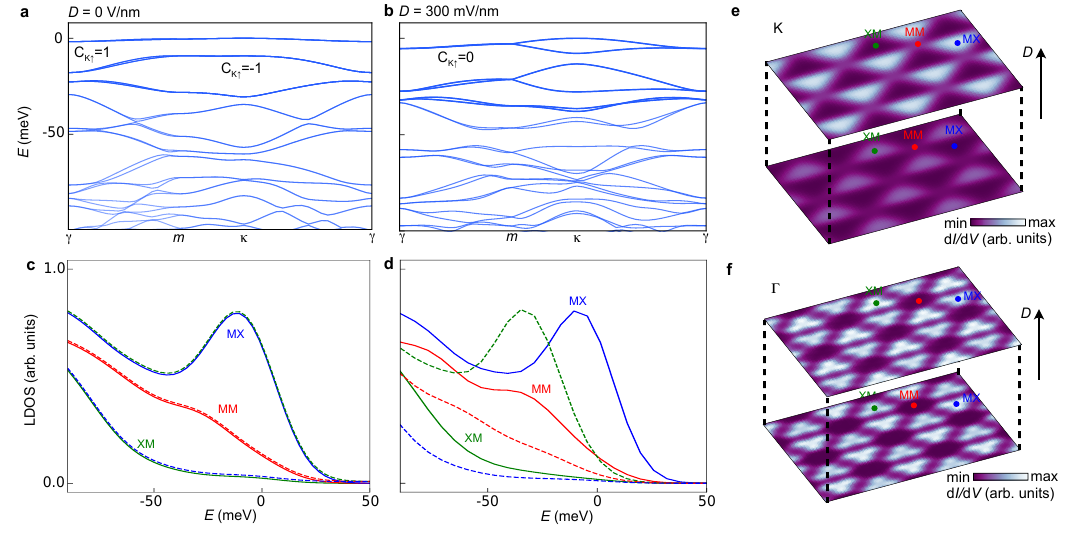}
\caption{\textbf{Predicted effects of displacement field on the K-point states.} 
\textbf{a-b}, Calculated bandstructure for tMoTe$_2$ with $\theta=2.88^{\circ}$ with \textbf{a}, $D=0~\mathrm{mV/nm}$, and \textbf{b}, $D=300~\mathrm{mV/nm}$. 
\textbf{c-d}, Calculated LDOS at different high symmetry points within the \moire unit cell. The top (bottom) layer is shown as the solid (dashed) curve. For $D=0$, the curves from the top and bottom layer are perfectly C$_{2x}$ symmetric (note that the curves from the bottom layer are offset by 0.0002 for visual clarity). For $D\neq0$, C$_{2x}$ symmetry is broken and there is an energy splitting between the LDOS of different stacking sites in the two layers. 
\textbf{e}, Calculated spatially resolved LDOS at $E=-10~\text{meV}$ for both the top and bottom MoTe$_2$ layers with $D=300~\mathrm{mV/nm}$. The top layer shows localization on MX, while the bottom shows localization on XM, similar to in Fig.~\ref{fig:4}d. However, in this calculation the overall value of the LDOS across the \moire is notably larger on the top layer as a consequence of the displacement field. 
\textbf{f}, Calculated spatially resolved LDOS at $E=-170~\text{meV}$ and $D=300~\mathrm{mV/nm}$, corresponding to states arising from the $\Gamma$ point. Because of their interlayer hybridization, these states depend only very weakly on $D$, and this calculation is nearly indistinguishable from the $D=0$ case shown in Fig.~\ref{fig:4}f. 
}
\label{fig:D_dependence}
\end{figure*}
%%%%%%%%%%%%%%%%%%%%%%%%%%%%%%%%%%%%%%%%%%%%%%%%%%%%%%%%%%%%%%%%%%%%%

%%%%%%%%%%%%%%%%%%%%%%%%%%%%%%%%%%%%%%%%%%%%%%%%%%%%%%%%%%%%%%%%%%%%%
\begin{figure*}
\includegraphics[width=\textwidth]{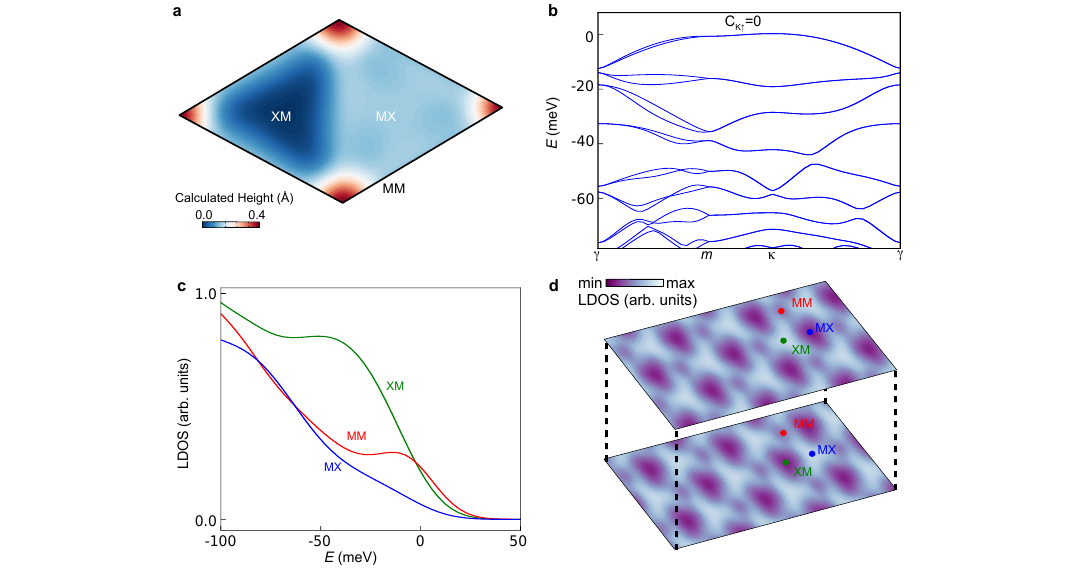}
\caption{\textbf{Calculated LDOS excluding the in-plane lattice relaxations.} 
\textbf{a}, Height of top layer in 2.88$^{\circ}$ tMoTe$_2$ predicted by an atomic relaxation DFT calculation excluding in-plane lattice relaxations (but retaining the out-of-plane relaxations). The height profile is identical to that shown in Fig.~\ref{fig:1}c.
\textbf{b}, Calculated band structure with in-plane lattice relaxations excluded. The first moir\'e valence band has a valley Chern number of 0, in contrast to its value of 1 when in-plane relaxations are included.
\textbf{c}, Corresponding calculations of LDOS at MM, MX, and XM in the top MoTe$_2$ layer as a function of energy near the valence band edge.
\textbf{d}, Layer-resolved LDOS at $-10$ meV, showing localization on the XM sites in the top layer and the MX sites in the bottom layer. This result is in contrast to the calculated LDOS including in-plane lattice relaxations shown in Fig.~\ref{fig:4}, in which the LDOS is localized on MX in the top layer near the band edge. 
}
\label{fig:no_relaxations}
\end{figure*}
%%%%%%%%%%%%%%%%%%%%%%%%%%%%%%%%%%%%%%%%%%%%%%%%%%%%%%%%%%%%%%%%%%%%%

\clearpage

\end{document}

%\linenumbers

\title{Supplementary Information: Visualizing the microscopic origins of topology in twisted molybdeum ditelluride}

\author{Ellis Thompson$^{1*}$}
\author{Keng Tou Chu$^{1*}$}
\author{Florie Mesple$^{1*}$}
\author{Xiao-Wei Zhang$^{2*}$}
\author{Chaowei Hu$^{1}$}
\author{Yuzhou Zhao$^{1,2}$}
\author{Heonjoon Park$^{1}$}
\author{Jiaqi Cai$^{1}$}
\author{Eric Anderson$^{1}$}
\author{Kenji Watanabe$^{3}$} 
\author{Takashi Taniguchi$^{4}$}
\author{Jihui Yang$^{2}$}
\author{Jiun-Haw Chu$^{1}$}
\author{Xiaodong Xu$^{1,2}$}
\author{Ting Cao$^{2\dagger}$}
\author{Di Xiao$^{2,1\dagger}$}
\author{Matthew Yankowitz$^{1,2\dagger}$}

\affiliation{ $^{1}$ Department of Physics, University of Washington, Seattle, Washington, 98195, USA}
\affiliation{$^{2}$ Department of Materials Science and Engineering\, University of Washington\, Seattle\, Washington\, 98195\, USA}
\affiliation{$^{3}$ Research Center for Functional Materials\, National Institute for Materials Science\, 1-1 Namiki\, Tsukuba 305-0044\, Japan}
\affiliation{$^{4}$ International Center for Materials Nanoarchitectonics\, National Institute for Materials Science\, 1-1 Namiki\, Tsukuba 305-0044\, Japan}
\affiliation{$^{*}$ These authors contributed equally to this work.}
\affiliation{$^{\dagger}$ $tingcao@uw.edu$ (T.C.); $dixiao@uw.edu$ (D.X.); $myank@uw.edu$ (M.Y.)}

\maketitle

\clearpage

\section{Constant-current spectroscopy}
\begin{figure*}
\includegraphics[width=\textwidth]{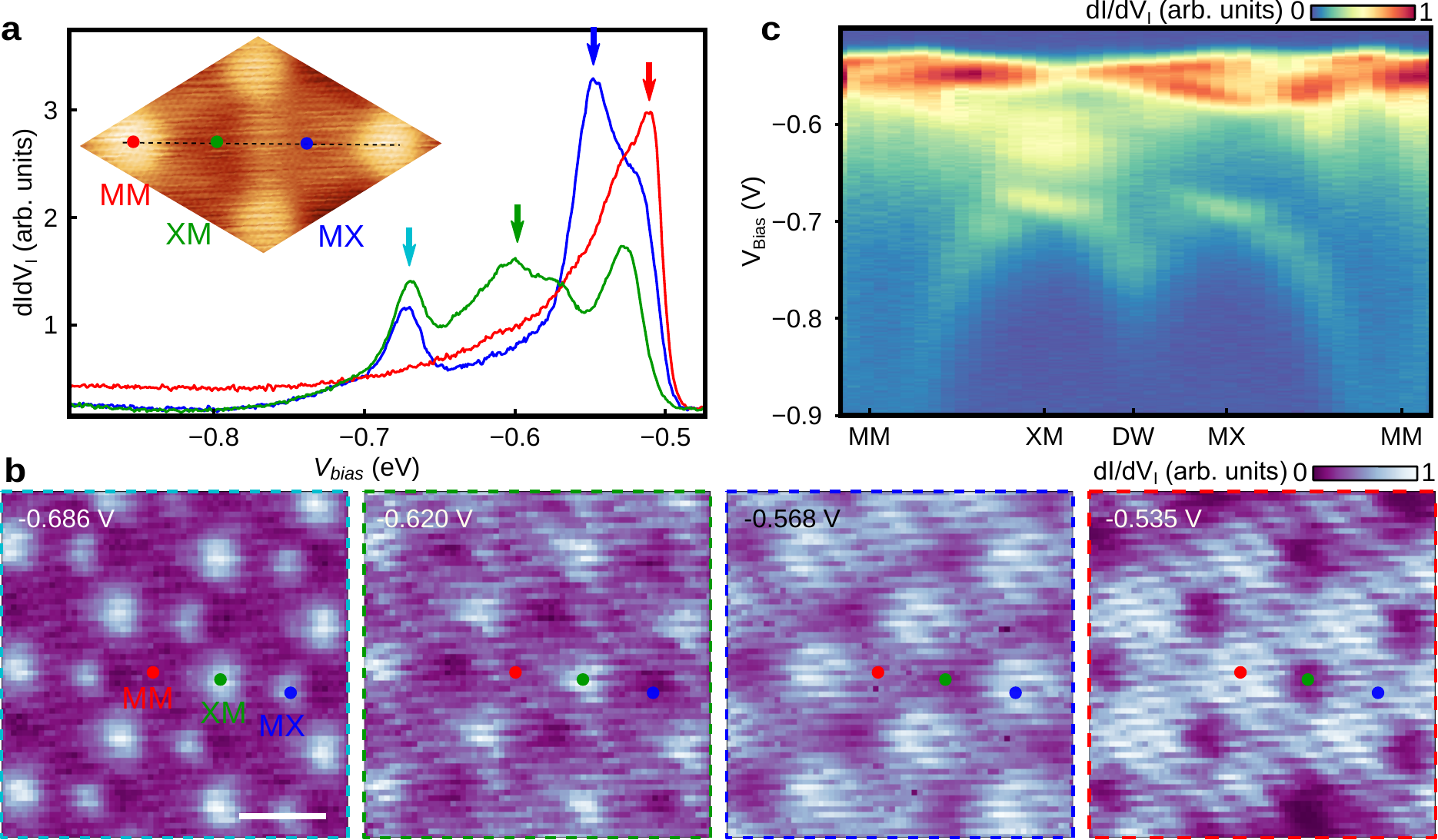} 
\caption{\textbf{Constant-current spectroscopy of the valence band edge.} 
\textbf{a}, Constant-current d$I$/d$V_I$ spectra acquired at different high symmetry points (2x2 pixels, 0.62~nm$^2$) for the same $\theta=2.75^{\circ}$ region shown in Figs. 1e, 3 and 4a-b of the main text. The highest intensity peaks correspond to the valence band edge, where the signal can have non-negligible contribution from the d$z/$d$V$ component that diverges when the tip-sample separation reduces in order to tunnel directly into the underlying graphene $p_z$ orbitals. This signal is thus likely to be very dependent on the relative height of the different tMote$_2$ regions, and could produce artifacts at the MM region owing to its large height. The inset shows the topography for this region, and a 10~nm long dashed line. 
\textbf{b}, Corresponding d$I$/d$V_I$ maps at various $V_{bias}$ showing spatial localization of LDOS at the MX/XM, XM, MX, and then MM regions, successively, from left to right. The origin of the localization on XM at $V_{bias}=-0.62$~V is currently unclear.
\textbf{c}, Line cut of the constant-current spectra along the dashed path from \textbf{a}, showing the smooth spatial variation of the band-edge localization. The arch-like features for $V_{bias}\lessapprox-0.67$~V correspond to $\Gamma$-point states, consistent with the behavior seen in Fig.~3b of the main text. Initial tunneling setpoint is $(I_t, V_{bias}) = (10~\text{pA}, -0.9~\text{V})$, scale bar is 5~nm.}
\label{fig:constant_current}
\end{figure*}

In addition to the constant-height spectroscopy techniques employed in the main text, we have also studied tMoTe$_2$ using constant-current spectroscopy. This is a dynamic spectroscopy measurement in which the current feedback loop is maintained during the entire $V_{bias}$ sweep, allowing the tip height to fluctuate with the current maintained at a constant value while a d$I$/d$V$ spectrum is acquired. Representative constant-current spectroscopy data is shown in Fig.~\ref{fig:constant_current}. 

Assuming the tip DOS is featureless in energy and operating at zero temperature, the tunneling current is a function of the sample LDOS, $\nu(E)$, and tunneling probability, $T(E,V,z)$ : 
\begin{align}
    I(V)\propto \int_{E_F}^{E_F+eV}\nu(E)T(E,V,z)dE, \hspace{12pt}T(E,V,z) = e^{-2\kappa(E,V)z}
\end{align}
where $\kappa$ is the decay constant, $z$ is the tip-sample distance, and $V$ is the tip-sample bias. It follows that several terms contribute to the measured differential tunneling conductance, d$I$/d$V$:
\begin{align}\label{eq:dIdV_cc}
\frac{\text{d}I}{\text{d}V} \propto 
\nu(E_F+eV)T(E_F+eV,V,z)
+\int_{E_F}^{E_F+eV}\nu(E)\frac{\text{d} T(E,V,z)}{\text{d} V}dE.
% \label{eq:dIdV}
\end{align}

With standard constant-height spectroscopy, the second term above is often neglected as $z$ remains constant and $\kappa$ is assumed to vary slowly over the bias sweep. With constant-current spectroscopy, however, $z$ is not fixed and is instead a function of $V$ set by the constraint $I=\text{const}$; still the d$I/$d$V$ signal can be attributed mostly to DOS~\cite{Zhang2015,Zhang2020} (i.e., the first term in Supplementary Information Eq.~(\ref{eq:dIdV_cc})) as long as the tip height and decay constant variations are sufficiently small. In a case where any of these conditions break down, this second term in Supplementary Information Eq.~(\ref{eq:dIdV_cc}) represents a non-negligible contribution the the measured d$I$/d$V$. This contribution can be separated into two terms dominated by the variation of $\kappa$ and $z$: 
\begin{align}\label{eq:dTdV}
\frac{\text{d} T(E,V,z)}{\text{d} V} = 
-2\bigg{[}\kappa(E,V)\frac{\text{d}z(V)}{\text{d}V}+z(V)\frac{\partial\kappa(E,V)}{\partial V}\bigg{]}
e^{-2\kappa(E,V)z(V)}.
\end{align} 

To examine the effect of these two terms, $\kappa(E,V)$ can first be estimated via the expression,
\begin{align}\label{eq:kappa}
    \kappa=\sqrt{\frac{2m(\phi_0-E-e\vert V\vert /2)}{\hbar^2}+k_{\vert \vert}^2(E)},
\end{align}
where $\phi_0$ is the average work function of the tip and the sample and $k_{\vert \vert}$ is the parallel momentum. Here, because the relevant work functions are a few times larger than the applied bias \cite{workfunction-Aftab,workfunction-Mleczko}, the variation of $\kappa$ with $V$ will be small and the second term in Supplementary Information Eq.~(\ref{eq:dTdV}) still gives negligible contribution. 
%In the case of tMoTe$_2$ on graphene, $z$ varies rapidly with $V$ near the MoTe$_2$ valence band edge since the dominant tunneling channel changes abruptly from the $d_{xy}+d_{x^2-y^2}$ orbitals from the MoTe$_2$ K-points to the $p_z$ orbitals from the underlying graphene.

However, as the bias is tuned out of the valence band and into the semiconducting gap of MoTe$_2$ and the dominant tunneling channel switches, the tip experiences a sudden drop in height towards the sample. As a consequence, the assumption of a slowly varying $z$ as a function of bias breaks down, and the first term in Supplementary Information Eq.~(\ref{eq:dTdV}) can become relevant as $\text{d}z(V)/\text{d}V$ is now a large negative quantity. This is particularly true at the MM regions of the tMoTe$_2$, where the distance between the top layer of MoTe$_2$ and the graphene is largest and a more pronounced drop in tip height is expected. This may explain the large peak in the constant-current spectra on MM denoted by the red arrow in Fig.~\ref{fig:constant_current}a, which is not seen in the reduced-height spectroscopy shown in Fig.~4b of the main text. 

In summary, the constant-current spectroscopy technique can enable a good identification of the energy of the band edge even when $\kappa$ is large, as in tMoTe$_2$. However, we also find that it can be prone to artifacts owing to the complexity of both physical and electronic structure of tMoTe$_2$. In particular, the moir\'e-scale corrugations of the tMoTe$_2$ can result in additional complications not present in flat monolayer TMDs~\cite{Zhang2015}. Therefore, a direct interpretation of the constant-current spectroscopy in the context of the LDOS requires careful consideration and modeling of $\kappa$ and $z$ as a function of energy and bias, which is beyond the scope of our study. 

%%%%%%%%%%%%%%%%%%%%%%%%%%%%%%%%%%%%%%%%%%%%%%%%%%%%%%%%%%%%%%%%%%%%%
%%%%%%%%%%%%%%%%%%%%%%%%%%%%%%%%%%%%%%%%%%%%%%%%%%%%%%%%%%%%%%%%%%%%%

\section{Reduced-height spectroscopy}

%%%%%%%%%%%%%%%%%%%%%%%%%%%%%%%%%%%%%%%%%%%%%%%%%%%%%%%%%%%%%%%%%%%%%
\begin{figure*}
\includegraphics[width=.9\textwidth]{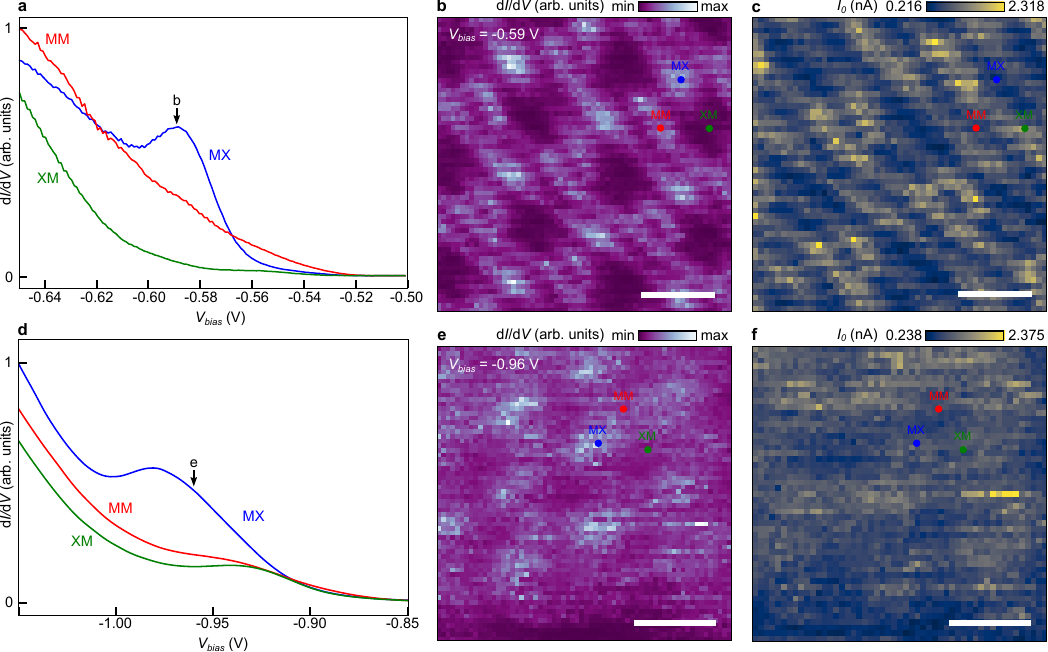}
\caption{\textbf{Unnormalized reduced-height spectroscopy}
\textbf{a}, Representative unnormalized reduced-height spectra averaged at three different high-symmetry sites on the $\theta=2.75^\circ$ region. 
\textbf{b}, d$I/$d$V$ map at the MX peak energy indicated by the arrow in \textbf{a}, obtained from the same set of reduced-height spectra. 
\textbf{c}, Initial current, $I_0$, of the reduced height spectra taken at each point in space. \textbf{a-c} are from the same dataset as Fig.~4a-b of the main text. 
\textbf{d-f}, Same as \textbf{a-c}, but for a $\theta=3.48^\circ$ region, from the same dataset as Extended Data Fig.~6. All scale bars are 5~nm.}
\label{fig:supp_variedz_raw}
\end{figure*}
%%%%%%%%%%%%%%%%%%%%%%%%%%%%%%%%%%%%%%%%%%%%%%%%%%%%%%%%%%%%%%%%%%%%%

\subsection{Normalization}
When performing reduced-height spectroscopy (which is fundamentally the same as constant-height spectroscopy), $z$ is kept constant with the feedback loop open. Because the relevant work functions are a few times larger than the applied bias \cite{workfunction-Mleczko,workfunction-Aftab}, the tunneling barrier and thus $\kappa$ is effectively $V$-independent according to Supplementary Information Eq.~(\ref{eq:kappa}). The second term in Supplementary Information Eq.~(\ref{eq:dIdV_cc}) can therefore be neglected. In the case of tunneling primarily into the $K$-point states of MoTe$_2$, which is true near the band edge, $\kappa$ can be further approximated to be $E$-independent. This gives
\begin{align}
    I \propto e^{-2\kappa z}\int_{E_F}^{E_F+eV}\nu(E)dE\implies
    \frac{\text{d}I}{\text{d}V} \propto e^{-2\kappa z}\nu(E_F+eV).
\end{align}

Although the d$I/$d$V$ signal is purely proportional to the LDOS here (as opposed to the case of constant-current spectroscopy described above), the proportionality factor can still vary from spectrum to spectrum due to slight differences in the stabilization condition leading to slightly different $z$. At small $z$, in particular, this effect is exponentially amplified. This artifact can be mitigated by normalizing the measured d$I/$d$V$ by the current at a specific bias, $I_0=I(V_0)$, which gives
\begin{align}\label{eq:normalized}
    \frac{1}{I_0}\frac{\text{d}I}{\text{d}V} \propto \frac{\nu(E_F+eV)}{\int_{E_F}^{E_F+eV_0}\nu(E)\text{d}E}.
\end{align}
This quantity is proportional to the DOS and does not depend on $z$, which makes the comparison between spectra taken at positions with similar $\nu(E)$ structure, i.e. in the vicinity of equivalent high-symmetry sites, more consistent. In our case, $I_0$ was chosen to be the initial current of each spectrum. However, since $\nu(E)$ varies across the \moire unit cell, the denominator in Supplementary Information Eq.~(\ref{eq:normalized}) can be highly position-dependent. Therefore, special care has to be taken when comparing normalized spectra across different points in the unit cell.

Fig.~\ref{fig:supp_variedz_raw}a(d) shows the unnormalized reduced-height d$I$/d$V$ spectra averaged at three different high-symmetry sites for tMoTe$_2$ with $\theta=2.75^{\circ}$ ($3.48^{\circ}$). The shapes of the spectra are largely the same as the normalized cases shown in Fig.~4b (Extended Data Fig.~6a), exhibiting a spectral peak only at the MX site. Fig.~\ref{fig:supp_variedz_raw}b(e) shows a d$I$/d$V$ map acquired at a value of $V_{bias}$ within the peak, assembled from the same unnormalized reduced-height spectra. Similarly, the map reflects the same hierarchy of localization at different stacking sites as shown in the normalized map in Fig.~4a (Extended Data Fig.~6e), albeit with noticeable noise. This noise is primarily a consequence of fluctuations in the stabilization condition of each spectrum, appearing also in the $I_0$ map shown in Fig.~\ref{fig:supp_variedz_raw}c(f). Normalizing by $I_0$ thus eliminates this noise, yielding the spectra and maps shown in Fig.~4a-b (Extended Data Fig.~6). Furthermore, $I_0$ shows minimal correlation to the real-space position at which the corresponding spectrum is taken. This demonstrates that the normalization procedure introduces minimal artifacts and instead simply removes a significant portion of the noise.

\subsection{Residual signal near the tMoTe$_2$ valence band edge}
The reduced-height spectra in Fig.~4a show a residual signal on MM for a small range of $V_{bias}$ below the MX peak. Extended Data Fig.~6h further shows this residual signal appears at both MM and the DWs (see also Supplementary Video 2). We attribute this mainly to position-dependent tip-induced doping caused by the out-of-plane buckling of the tMoTe$_2$ layers. There may be more tip-induced electron doping when the top MoTe$_2$ layer is furthest from the grounded graphene, causing the the band edge to shift towards $V=0$ at regions with larger interlayer spacing (i.e., MM and the DWs). Additionally, tip-induced deformations of the layers can cause an increase in the tunneling current or create additional inelastic tunneling channels which might offset the measured d$I/$d$V$ signal. Closer to $V_{bias}=0$, a faint residual signal is visible on XM, which is shown in Supplementary Video 2. This can be explained by enhanced tunneling into the underlying graphene at XM, where the tip is physically closest to the graphene. 

\section{Effects of twist angle and strain on the localization of $\Gamma$-point states}
\label{sec:Gamma_dependency}

%%%%%%%%%%%%%%%%%%%%%%%%%%%%%%%%%%%%%%%%%%%%%%%%%%%%%%%%%%%%%%%%%%%%%
\begin{figure*}
\includegraphics[width=\textwidth]{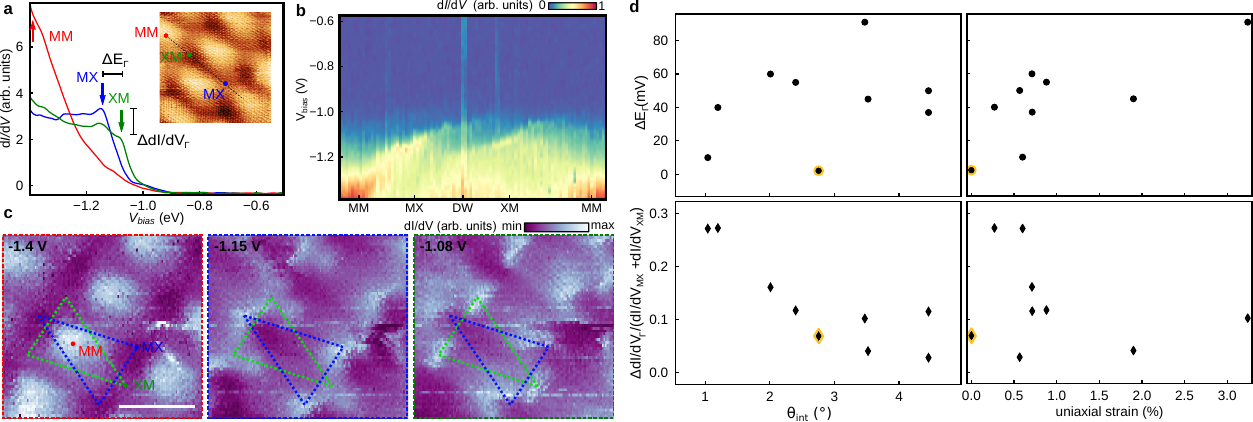} 
\caption{\textbf{Dependence of the localization of $\boldsymbol{\Gamma}$ states on the local stacking configuration.}
\textbf{a}, Constant-height tunneling spectra acquired at different high symmetry points for a highly strained region with stacking configuration $\theta = 3.47^{\circ}$, $\varepsilon_{uni} = 3.24\%$ oriented at 34$^\circ$ with respect to the moir\'e, $\varepsilon_{bi}=-0.73\%$. 
\textbf{b}, Line cut of the constant-height spectra acquired along the 10~nm long dashed path in \textbf{a}, showing shifting of the $\Gamma$ band edge within the MX/XM regions. 
\textbf{c}, d$I$/d$V$ maps acquired at different $V_{bias}$ values as indicated by the arrows in \textbf{a}, showing an evolution of LDOS localization from MM regions to MX/XM  regions. Localization in MX/XM regions is deformed uniaxially along the direction of the heterostrain. All data in this figure is assembled from a grid spectroscopy measurement with initial tunneling setpoint ($I, V_{bias}$) = ($300$~pA, $-1.4$~V). Scale bar is 5~nm.
\textbf{d}, Systematic study of the energy splitting, $\Delta E_\Gamma$, and the amplitude ratio between the d$I$/d$V$ peaks at MX and XM, plotted as a function of twist angle and uniaxial heterostrain, $\varepsilon_{uni}$. We find that the amplitude ratio is best correlated with $\theta$, whereas the energy splitting with $\varepsilon_{uni}$. The points highlighted in yellow correspond to the region shown in Figs.~2-4 of the main text.
}
\label{fig:SupplStrainedRegion}
\end{figure*}
%%%%%%%%%%%%%%%%%%%%%%%%%%%%%%%%%%%%%%%%%%%%%%%%%%%%%%%%%%%%%%%%%%%%%

We have measured the localization of $\Gamma$-point states using constant-height spectroscopy at several regions of tMoTe$_2$ with varying twist angle and uniaxial strain, $\varepsilon_{uni}$, which both modify the local atomic stacking arrangements within the \moire lattice. We find that that the d$I$/d$V$ peak from the edge of the $\Gamma$-point bands is always localized on the XM and MX regions. We additionally find that this localization pattern often breaks C$_{2x}$ symmetry, such that the amplitude and bias of the d$I$/d$V$ peak differ between the XM and MX regions. The magnitude of this symmetry breaking varies widely for different stacking configurations. This is weakly visible in the region with $\theta=2.75^{\circ}$ in Fig.~3 of the main text, where the $\Gamma$-peak at XM has a slightly larger amplitude than at MX. In contrast, the region with $\theta=3.52^{\circ}$ shown in Extended Data Fig.~3 features a larger $\Gamma$ peak on MX. In addition, the two peaks are split in bias by $\approx50$~mV in this region.

We also observe complex spatial localization structures of the $\Gamma$-point feature within the MX and XM regions, visible in the localization patterns at intermediate energies shown in Fig.~3c of the main text. This is further exemplified by measurements of a region with very large strain, Fig.~\ref{fig:SupplStrainedRegion}a, with stacking arrangement $\theta=3.47^{\circ}$, $\varepsilon_{uni}=3.24\%$ and $\varepsilon_{bi}=-0.73\%$. In this case, the $\Gamma$-point states in the MX and XM region appear as shoulder-like rather than as peaks in d$I$/d$V$, and differ both in energy and amplitude between the sites (similar to the data shown in Extended Data Fig.~3). Fig.~\ref{fig:SupplStrainedRegion}b shows a line cut of constant height spectra along a path that crosses the MM, MX, and XM regions, indicated by the black dashed line in the inset of panel a. This map shows that the $\Gamma$-point states shift rapidly in $V_{bias}$ across the MX and XM regions. Spatial maps of d$I$/d$V$ acquired at various fixed $V_{bias}$ (Fig.~3c) further show the LDOS localization along strained elongated bright regions surrounding MX and XM. These sharply contrast the maps shown in Fig.~3c of the main text which form an almost perfectly symmetric hexagonal structure owing to the reduced uniaxial strain. 

\begin{figure*}
\includegraphics[width=\textwidth]{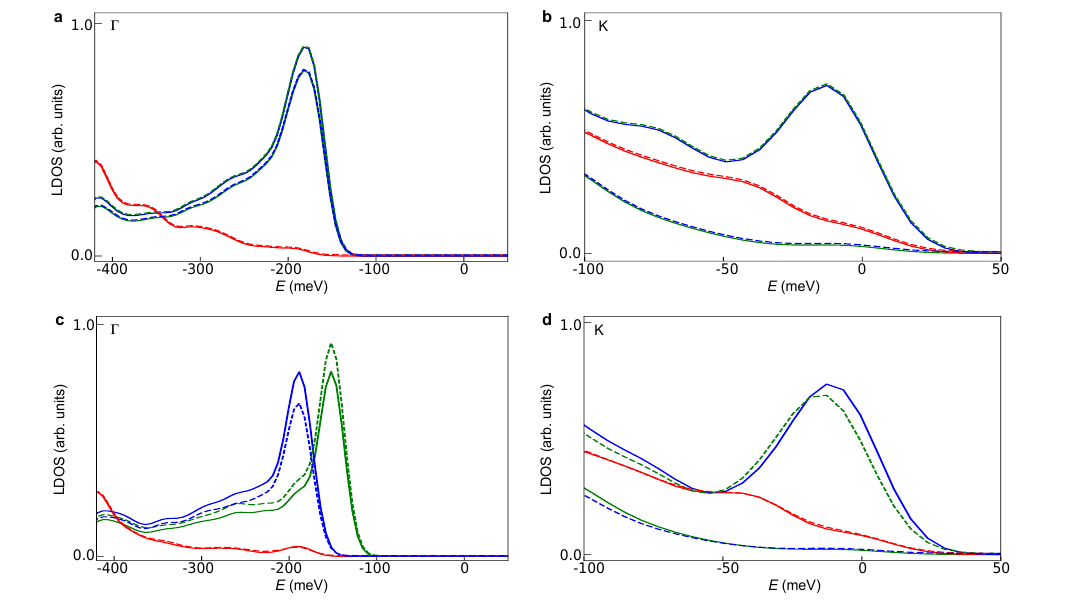} 
\caption{\textbf{Effect of the interlayer distance on the calculated LDOS.} 
\textbf{a-b}, Calculated LDOS at different stacking sites within the \moire unit cell for $\theta=3.48^{\circ}$ in the top layer (solid lines) and bottom layer (dashed lines). The calculation in \textbf{a} is for $\Gamma$-point states, and for \textbf{b} it is for K-point states.
\textbf{c-d}, Similar calculations but with the interlayer separation artificially increased by .05 $\mathrm{\AA}$ at the MX region and decreased by .05 $\mathrm{\AA}$ at the XM region. This difference in interlayer separation clearly induces a splitting between $\Gamma$-point peaks at the MX and XM regions, but has negligible effect on the K-point states. 
}
\label{fig:interlayer_distance}
\end{figure*}

Collectively, these observations suggest that localization of the $\Gamma$-point states depends sensitively on twist angle and strain. In Fig.~\ref{fig:SupplStrainedRegion}d, we plot the energy difference, $\Delta E_\Gamma$, and normalized amplitude ratio, $\frac{\Delta dI/dV_{\Gamma}}{dI/dV_{MX}+dI/dV_{XM}}$, between the $\Gamma$-point features at MX and XM for a variety of regions of tMoTe$_2$. The amplitude ratio appears to correlate better with $\theta$, whereas the energy splitting correlates better with $\varepsilon_{uni}$. However, neither show a definitive trend, possibly due to the large number of intertwined factors that can affect the localization of the $\Gamma$-point bands. One especially important parameter is the local interlayer separation, which can selectively shift the edge of $\Gamma$ bands in TMD heterobilayers~\cite{waters_flat_2020,olin_ab-initio_2023,brzezinska_pressure-tuned_2024}. This may explain why the $\Gamma$-point features shift significantly from the MX/XM regions to the domain walls to the MM regions, which all have notably different interlayer separation based on the DFT calculations shown in Fig.~1c of the main text. Although these calculations predict a similar interlayer separation for MX and XM, they do not account for the presence of the graphene substrate or weak tip-sample interactions that can change the relative interlayer spacing. For instance, the XM region is closer to the graphene than the MX, which may have the effect of slightly changing the XM spacing compared to MX. Figure~\ref{fig:interlayer_distance} shows our calculations of the LDOS at different high symmetry points for different values of the interlayer separation at the MX and XM regions. We see that a splitting in the energy of the $\Gamma$-point band edge appears when the interlayer spacing differs between MX and XM, pointing to this effect as a plausible explanation for our observations in strained tMoTe$_2$. Additionally, other electrostatic factors, such at a large electric field from the tip and the effect of the graphene isopotential, may lead to slight differences between the magnitude of the LDOS between the MX and XM regions (see Extended Data Fig.~8).

%%%%%%%%%%%%%%%%%%%%%%%%%%%%%%%%%%%%%%%%%%%%%%%%%%%%%%%%%%%%%%%%%%%%%
%%%%%%%%%%%%%%%%%%%%%%%%%%%%%%%%%%%%%%%%%%%%%%%%%%%%%%%%%%%%%%%%%%%%%

\section{Tip-Sample Interaction}

Probing the \moire bands at the valence band edge, which originate from the K point of tMoTe$_2$, requires a reduced tip-sample distance due to the large decay constant~\cite{Zhang2015,Zhang2020} as discussed in the main text. As a consequence, the tip may physically interact strongly with the MoTe$_2$ layers to cause a pressure-induced lattice commensuration that physically expands the energetically favorable stacking configurations and shrinks those that are disfavored. Similar phenomena have been observed in graphene/hBN \moire lattices~\cite{Yankowitz2016} and in twisted bilayer graphene (TBLG)~\cite{Choi2019,mesple2021}. Although it is micro-tip dependent, this effect often produces abrupt height jumps in the topography at regions of the moir\'e that have less stable atomic stacking configuration or higher local strain, such as at domain walls. In addition, it may create attendant artifacts in the d$I$/d$V$ spectroscopy maps.

\begin{figure*}
\includegraphics[width=.7\textwidth]{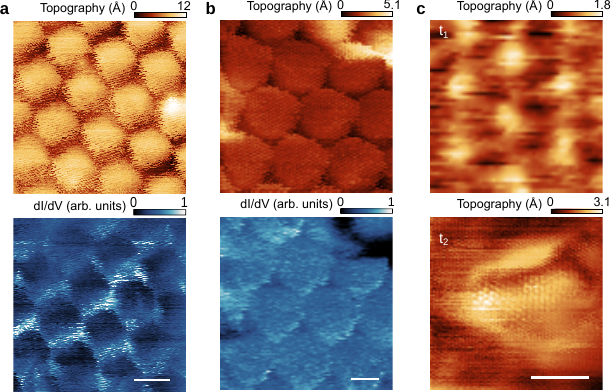} 
\caption{\textbf{Signatures of tip-sample interaction effects.}
\textbf{a-b}, Low-bias imaging of tMoTe$_2$ with $\theta=3.68^{\circ}$ and setpoints ($I_t$,$V_b$) of \textbf{a}, ($10$~pA,$-0.9$~V) and \textbf{b}, ($10$~pA,$-0.915$~V). These tunneling parameters correspond to small tip-sample separations, and we find that in these cases the tip directly modifies the local stacking of the layers and induces strong hexagonal reconstruction of the moir\'e lattice. 
\textbf{c}, Topography maps of the same region both before ($t_1$, top row) and after ($t_2$, bottom row) performing a series of low-bias LDOS maps. We often notice either a degradation (as shown here), or a local doping of the imaged tMoTe$_2$ region. Scale bars are 5~nm.}
\label{fig:supp_tip_interaction}
\end{figure*}

In the case of our tMoTe$_2$ measurements, this tip interaction effect, when present, appears to shrink the MM and XM regions and expand the MX regions. Figs.~\ref{fig:supp_tip_interaction}a-b show two examples of these tip-sample interaction effects in topography and constant-current d$I$/d$V$ maps. Both topography maps show deformed hexagonal \moire lattices with abrupt jumps at the domain walls. There is also artificially large d$I$/d$V$ signal around the locations of these jumps. Though these effects are similar to those observed previously in graphitic systems, they appear at much larger twist angles, reflecting the relative softness of the MoTe$_2$ membrane~\cite{shaffique2023}. Additionally, Fig.~\ref{fig:supp_tip_interaction}c shows the condition of the surface before (top) and after (bottom) performing a set of constant-current d$I$/d$V$ maps. There is a clear degradation of the surface after scanning at low bias while having the tip close to the surface. This is in contrast to reduced-height or constant-current grid spectroscopy maps, in which the scanning is performed at high bias and the tip only sporadically moves very close to the sample surface. We are therefore careful to avoid studying tMoTe$_2$ when these tip-sample interaction effects are present.

%%%%%%%%%%%%%%%%%%%%%%%%%%%%%%%%%%%%%%%%%%%%%%%%%%%%%%%%%%%%%%%%%%%%%
%%%%%%%%%%%%%%%%%%%%%%%%%%%%%%%%%%%%%%%%%%%%%%%%%%%%%%%%%%%%%%%%%%%%%

\clearpage

\bibliographystyle{naturemag}
\bibliography{references_2}